\documentclass[a4paper,12pt]{article}

\usepackage{latexsym,amsmath,amsfonts,amssymb}
\usepackage{multirow}
\usepackage{float,placeins}
\usepackage{graphicx,color}
\usepackage[FIGBOTCAP]{subfigure}
\usepackage{pstricks,pst-node,pst-text,pst-3d}
\usepackage{verbatim}

\newcommand{\Tr}{{\rm Tr}}

\begin{document}

\author{Simon Catterall}
\author{Luigi Del Debbio}
\author{Joel Giedt}
\author{Liam Keegan}
\title{MCRG Minimal Walking Technicolor}

\begin{titlepage}

\setcounter{page}{0}

\begin{flushright}
\end{flushright}

\vspace{0.6cm}

\begin{center}
{\Large \bf MCRG Minimal Walking Technicolor} \\  

\vskip 0.8cm

{\bf Simon Catterall}\\
{\sl Department of Physics, Syracuse University, Syracuse, NY 13244, USA}\\
{\bf Luigi Del Debbio}\\
{\sl The Tait Institute, School of Physics and Astronomy, The University of Edinburgh, Edinburgh EH9 3JZ, Scotland, UK}\\
{\bf Joel Giedt}\\
{\sl Department of Physics, Applied Physics and Astronomy, Rensselaer Polytechnic Institute, Troy, New York, 12180 USA}\\
{\bf Liam Keegan}\\
{\sl The Tait Institute, School of Physics and Astronomy, The University of Edinburgh, Edinburgh EH9 3JZ, Scotland, UK}\\

\vskip 1.2cm

\end{center}

\begin{abstract}
We present a Monte Carlo renormalisation group study of the SU(2) gauge theory with two Dirac fermions in the adjoint representation. Using the two--lattice matching technique we measure the running of the coupling and the anomalous mass dimension. We find slow running of the coupling, compatible with an infrared fixed point. Assuming this running is negligible we find a vanishing anomalous dimension, $\gamma=-0.03(13)$, however taking this source of systematic error into account gives a much larger range of allowed values, $-0.6 \lesssim \gamma \lesssim 0.6$. We also attempt to measure the anomalous mass dimension using the stability matrix method. We discuss the systematic errors affecting the current analysis and possible improvements.
\end{abstract}

\vfill

\end{titlepage}

\section{Introduction}

Technicolor theories with fermions in higher representations of the gauge group can potentially provide a dynamical electroweak symmetry breaking mechanism without conflicting with electroweak precision data. Minimal Walking Technicolor is an example of such a theory, a SU(2) gauge theory with two Dirac fermions in the adjoint representation~\cite{Sannino:2004qp,Luty:2004ye}. It is expected from perturbation theory to be in or near to the conformal window, although any new infrared fixed point (IRFP) is thought to occur at strong coupling and so non--perturbative results are necessary. Initial lattice simulations showed some evidence of walking dynamics and mapped out the phase diagram of the theory~\cite{Catterall:2007yx,Catterall:2008qk}. Subsequent Schr\"odinger Functional lattice simulations have indeed found that the gauge coupling runs very slowly~\cite{Bursa:2009we,DelDebbio:2009fd,Hietanen:2009az,DeGrand:2011qd}, more slowly than the perturbative prediction, and a recent simulation using Creutz ratios found evidence for backwards running at strong coupling~\cite{Giedt:2011kz}, although distinguishing between conformal and near--conformal behaviour is an inherently difficult task. Moreover, in order to be phenomenologically viable the theory must have a large anomalous mass dimension $(\gamma\sim1)$~\cite{Holdom:1981rm,Yamawaki:1985zg,Appelquist:1986an}, and recent work suggests that $\gamma>1$ is required~\cite{Chivukula:2010tn}.

A conjectured all--order beta function~\cite{Pica:2010mt} predicts $\gamma=11/24\simeq0.458$ for this model.\footnote{This prediction supersedes the
original all--order conjecture~\cite{Ryttov:2007cx} of $\gamma=3/4$ for this model.} This value is also consistent with perturbative results in the $\overline{\mathrm{MS}}$--scheme up to four loops~\cite{Pica:2010xq}.
The anomalous mass dimension has been measured non--perturbatively in recent lattice studies~\cite{Bursa:2009we,DelDebbio:2010hx,DeGrand:2011qd}, which have all found lower values.

In this work we measure the anomalous mass dimension using the Monte Carlo Renormalisation Group (MCRG) two--lattice matching method, a technique which has recently been used to investigate theories with many flavours of fermions in the fundamental representation~\cite{Hasenfratz:2009ea,Hasenfratz:2010fi}. We also investigate using the stability matrix MCRG method~\cite{Swendsen:1979gn}, which in principle allows the determination of all the critical exponents of a system. We consider the evolution of all possible couplings of the system under Wilson Renormalisation Group (RG) block transformations, where with each blocking step ultraviolet (UV) fluctuations are integrated out. Fixed points are characterised by the number of relevant couplings, which have positive scaling dimensions and flow away from the fixed point. Irrelevant couplings have negative scaling dimensions, and flow towards the fixed point, so that their infrared (IR) values are independent of their UV values.

\section{Two--Lattice Matching Method}

With each RG step, changing the scale by a factor $s$, irrelevant couplings will flow towards the fixed point (FP), and relevant couplings will flow away from it. After a few steps the irrelevant couplings should die out, leaving the flow following the unique renormalised trajectory (RT). If we can identify two sets of couplings which end up at the same point along the RT after the same number of steps, then they must have the same lattice correlation lengths, $\hat{\xi}=\hat{\xi}'$. Since the physical correlation length $\xi = \hat{\xi} a$ should not be changed by the RG transform, this means that they both must have the same lattice spacing $a$, or inverse cutoff $\Lambda^{-1} \sim a$. If they end up at the same point, but one takes an extra step, then their lattice correlation lengths must differ by a factor $s$, and hence so must their UV cutoffs.

To identify such a pair of couplings, we need to show that after $n$ and $(n-1)$ RG steps respectively their actions are identical. Explicitly calculating the actions would be complicated, but instead the gauge configurations themselves can be RG block transformed. Showing that the expectation values of all observables on these gauge configurations agree is equivalent to directly comparing the actions that generated them.

\label{sec:2lattice}
Starting with the SU(2) pure gauge theory, where the gauge coupling is the only relevant parameter, the procedure is as follows.

\begin{enumerate}
 \item Generate an ensemble of gauge configurations with an action $S(\beta)$ on a $L^4$ lattice.
 \item Block these $n$ times to produce an ensemble of configurations on a $(L/s^n)^4$ lattice, and measure the expectation values of various observables on them.
 \item Generate a new ensemble of gauge configurations with an action $S(\beta')$ on a $(L/s)^4$ lattice, for a range of values of $\beta'$.
 \item Block each of these $n-1$ times to produce an ensemble of configurations on a $(L/s^n)^4$ lattice, and measure the same observables for each $\beta'$.
 \item Interpolate in $\beta'$ such that each observable after taking $n$ steps on the larger lattice agrees with the same observable after taking $(n-1)$ steps on the smaller lattice.
 \item Repeat for different $n$, e.g. for $s=2,L=32$, three values can be used: $n=2,3,4$.
\end{enumerate}

\begin{figure}[ht]
  \centering
\includegraphics[angle=0,width=12cm]{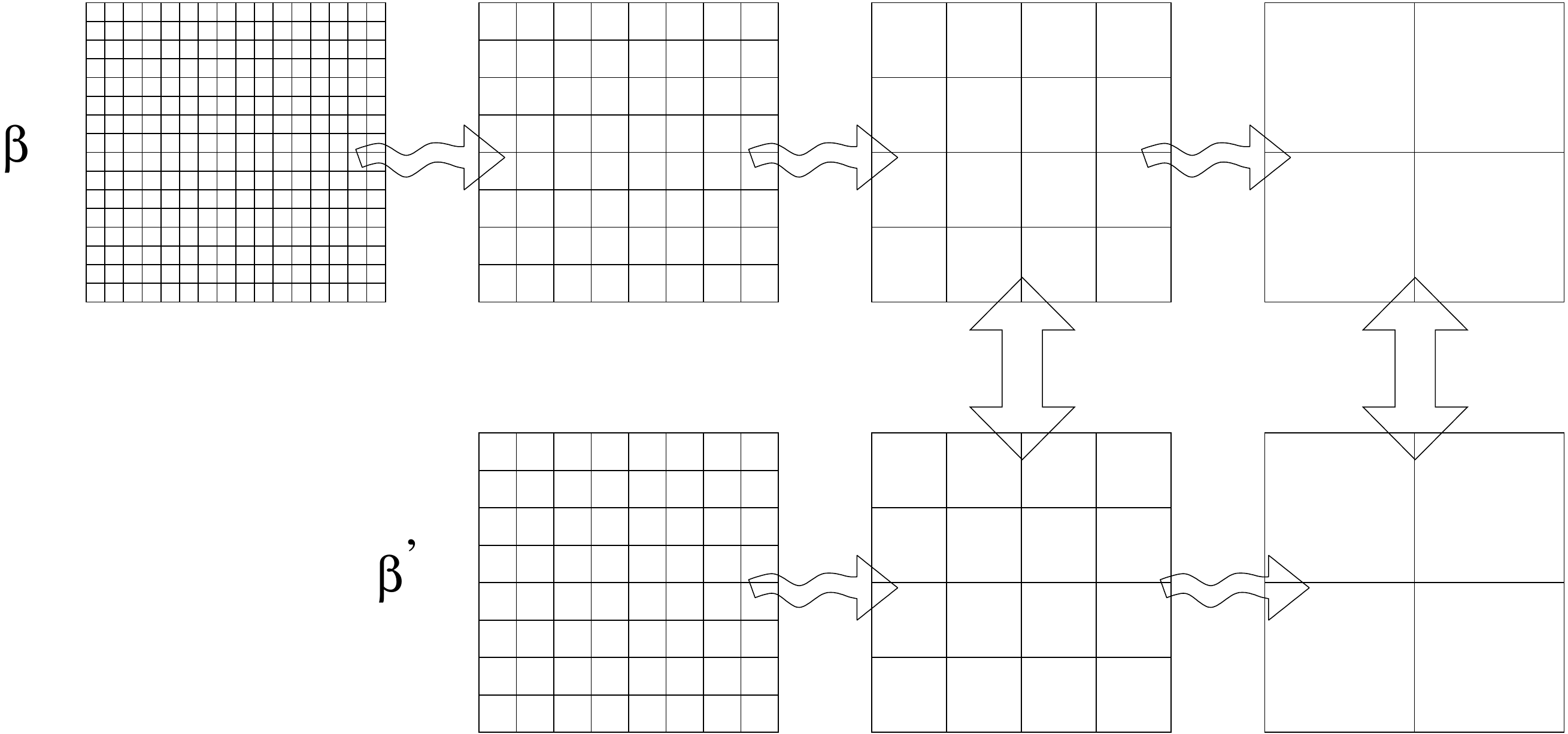}
  \caption{The two--lattice matching procedure described in Sec.~\ref{sec:2lattice}. Horizontal arrows represent RG blocking steps, and vertical arrows indicate matched lattices. In this case the matching is done after $2(1)$ and $3(2)$ blocking steps on the $16^4(8^4)$ lattices.}
  \label{fig:2lattice}
\end{figure}

We have now identified, for each $n$, a pair of bare gauge couplings $(\beta,\beta')$, with lattice correlation lengths that differ by a factor $s$, $\hat{\xi}' = \hat{\xi}/s$. In the limit $n \rightarrow \infty$, the quantity 
\begin{equation}
\Delta \beta = \beta - \beta' \equiv s_{b}(\beta;s)
\end{equation}
is the step scaling function for the bare gauge coupling. This is the analog of the Schr\"odinger Functional step scaling function for the renormalised coupling, $\sigma(u,s)$, and in the UV limit where $\overline{g}^2 \rightarrow g_0^2 = 2N/\beta$, there is a simple relation between the two:
\begin{equation}
\frac{s_{b}(\beta;s)}{\beta} = \frac{\sigma(u,s)}{u} - 1.
\end{equation}

We use the following $s=2$ RG blocking transform, labelled ORIG following the naming convention used in Ref.~\cite{Hasenfratz:2009ea}.
\begin{equation}
V^{\mathrm{ORIG}}_{n,\mu} = Proj\left[(1-\alpha)U_{2n,\mu}U_{2n+\mu,\mu}+\frac{\alpha}{6}\sum_{\nu\neq\mu}U_{2n,\nu}U_{2n+\nu,\mu}U_{2n+\mu+\nu,\mu}U_{2n+2\mu,\nu}^{\dagger}\right]
\end{equation}
where $U$ is the original gauge field on a $L^4$ lattice, $V$ is the blocked gauge field on a $(L/2)^4$ lattice, $Proj$ is the projection back into the group of SU(2) matrices, and $\alpha$ is a free parameter, which can be varied to optimise the transformation. Changing $\alpha$ changes the location of the FP, and how quickly the RT is approached in a given number of steps. Ideally it should be chosen such that

\begin{itemize}
 \item All operators predict the same matching coupling between $(n,n-1)$ pairs for a given blocking step $n$. (Deviations are a measure of the systematic error from not being at exactly the same point along the RT.)
 \item Consecutive blocking steps predict the same matching coupling, i.e. the coupling for which $(n,n-1)$ pairs agree should be the same for all $n$. (Deviations show that the RT is still being approached in the irrelevant directions\footnote{If there existed other relevant couplings that had not been precisely tuned to their critical values, these deviations could also indicate an unwanted flow in these relevant directions}.)
\end{itemize}

We also use two different $s=2$ RG blocking transforms constructed using hypercube (HYP) smeared links~\cite{Hasenfratz:2001hp}, labelled HYP and HYP2~\cite{Hasenfratz:2010fi}. HYP blocking uses a product of HYP smeared links with smearing parameters $(\alpha,0.6,0.3)$,
\begin{equation}
V^{\mathrm{HYP}}_{n,\mu} = W[U]_{2n,\mu}W[U]_{2n+\mu,\mu}
\end{equation}
where $U$ is the original gauge field and $W[U]_{x,\mu}$ is the HYP smeared link at $(x,\mu)$, defined in App.~\ref{app:hyp}.

HYP2 blocking is the same except that HYP smearing is applied twice to each link with smearing parameters $(\alpha,0.3,0.3)$.
\begin{equation}
V^{\mathrm{HYP2}}_{n,\mu} = W^2[U]_{2n,\mu}W^2[U]_{2n+\mu,\mu}
\end{equation}

Ideally our results would be independent of the choice of blocking, so the use of three different blocking transforms allows us to check the systematic errors of the procedure and the distance from the RT. The HYP and HYP2 blocking transforms have also been empirically found to work better than the ORIG transform at strong coupling~\cite{Hasenfratz:2010fi}.

\section{Pure Gauge Results}

\begin{figure}[ht]
  \centering
  \subfigure[$P(\mu,\nu)$]{\includegraphics[angle=270,scale=0.3]{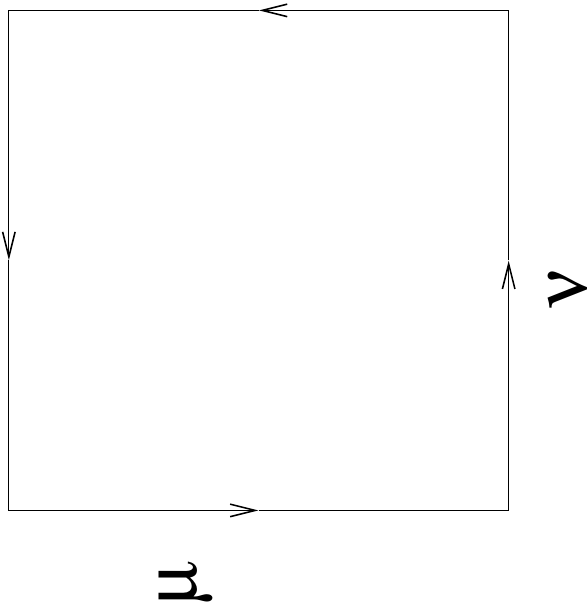}}\hspace{1.5cm}
  \subfigure[$L_6(\mu,\nu,\rho)$]{\includegraphics[angle=270,scale=0.3]{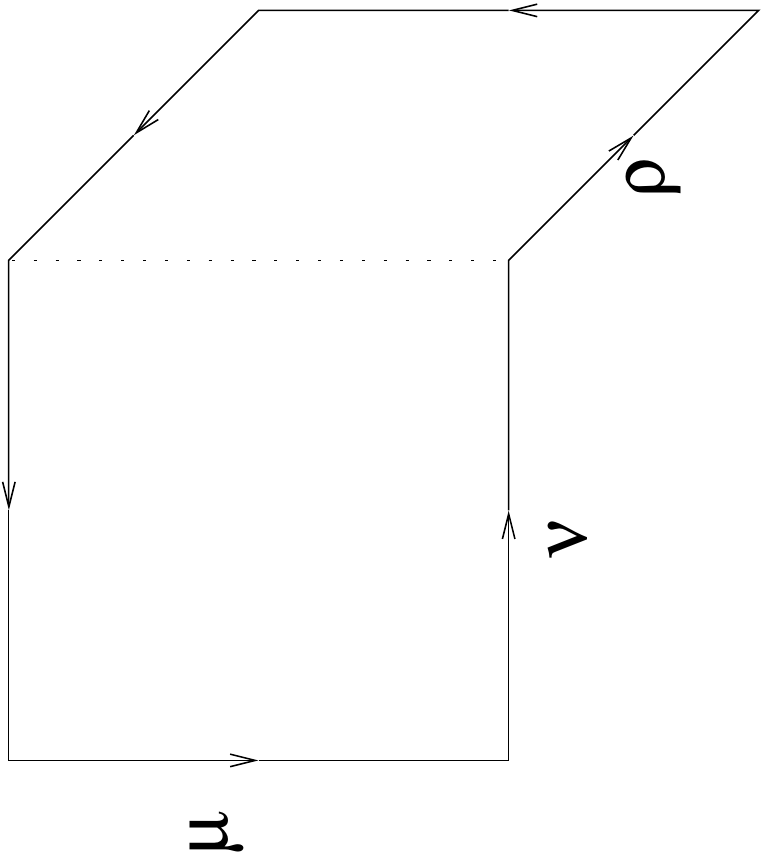}}\hspace{1.5cm}
  \subfigure[$L_8(\mu,\nu,\rho,\alpha)$]{\includegraphics[angle=270,scale=0.3]{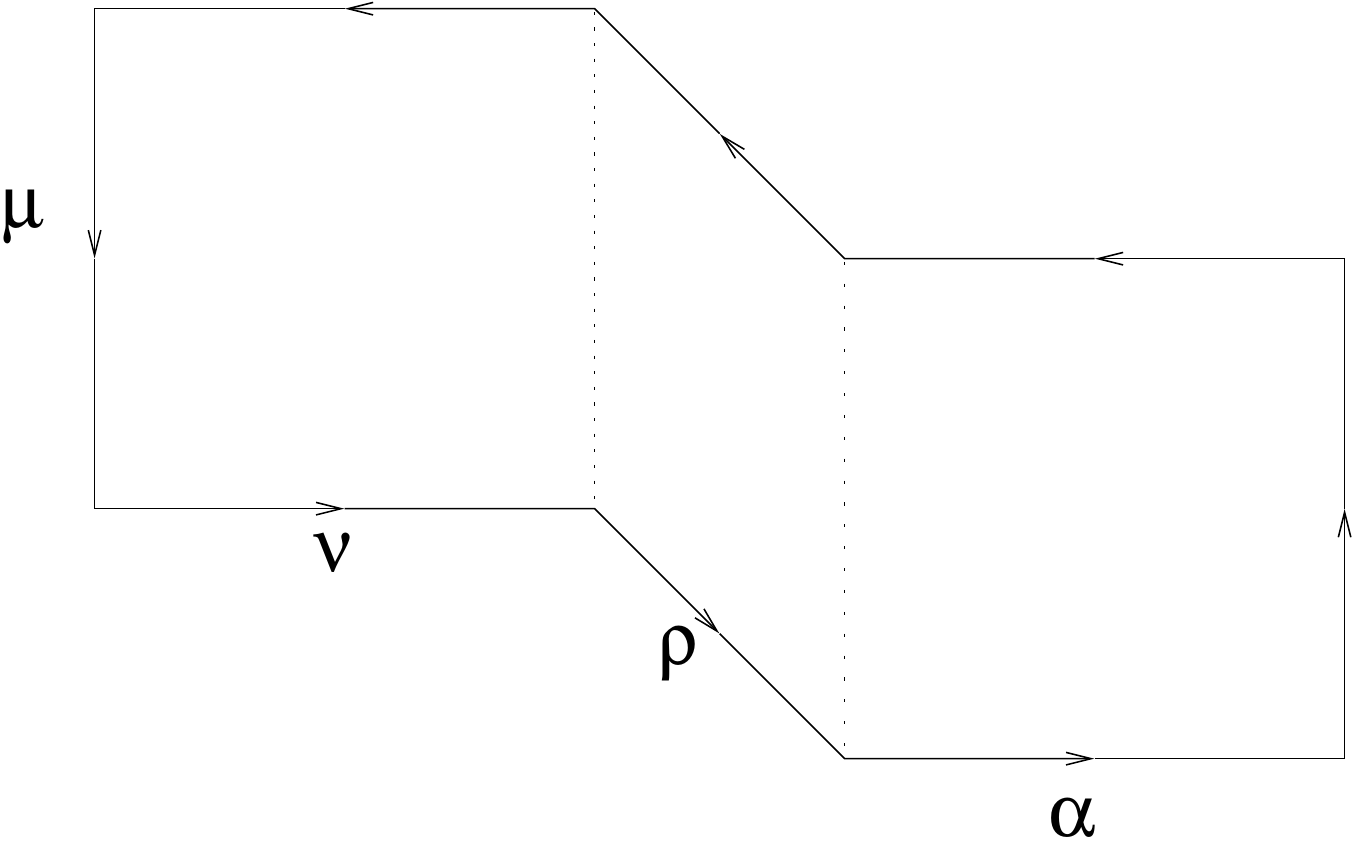}}
  \caption{The plaquette, 6--link and 8--link gauge observables used. Explicit definitions are given in App.~\ref{app:obs}.}
  \label{fig:PURE_loops}
\end{figure}

As an initial test of the method, matching in $\beta$ between 32(16) and 16(8) lattices was performed using $\sim2000$ SU(2) pure gauge configurations for each $\beta$. We matched in the plaquette, the three six--link loops, and three 8--link loops, shown in Fig.~\ref{fig:PURE_loops}. See App.~\ref{app:obs} for details of these observables.

\begin{figure}[ht]
  \centering
\subfigure[Plaquette Matching]{\label{fig:PURE_plaq_a}\includegraphics[angle=270,width=7.5cm]{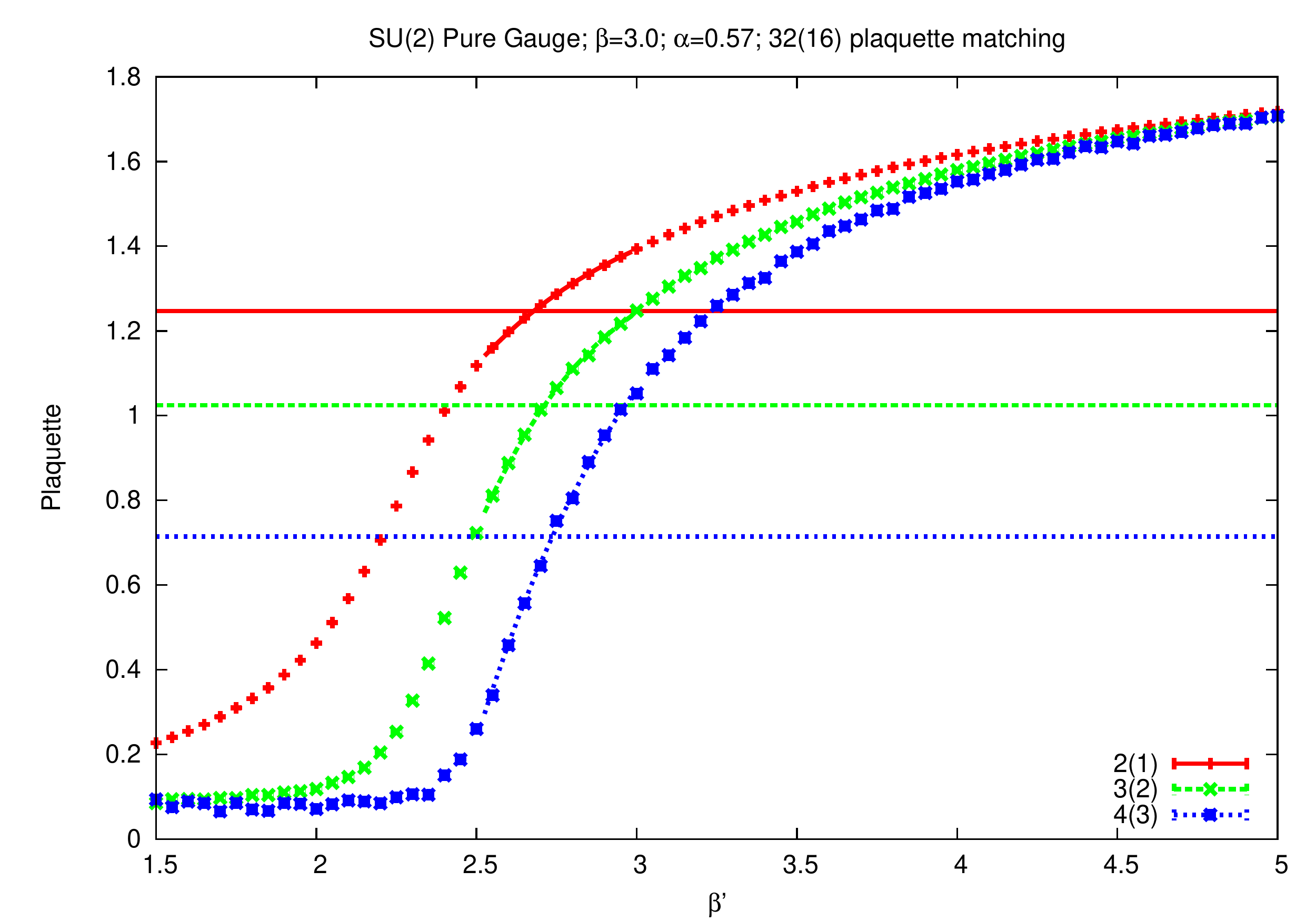}}
\subfigure[$\alpha$--Optimisation]{\label{fig:PURE_plaq_b}\includegraphics[angle=270,width=7.5cm]{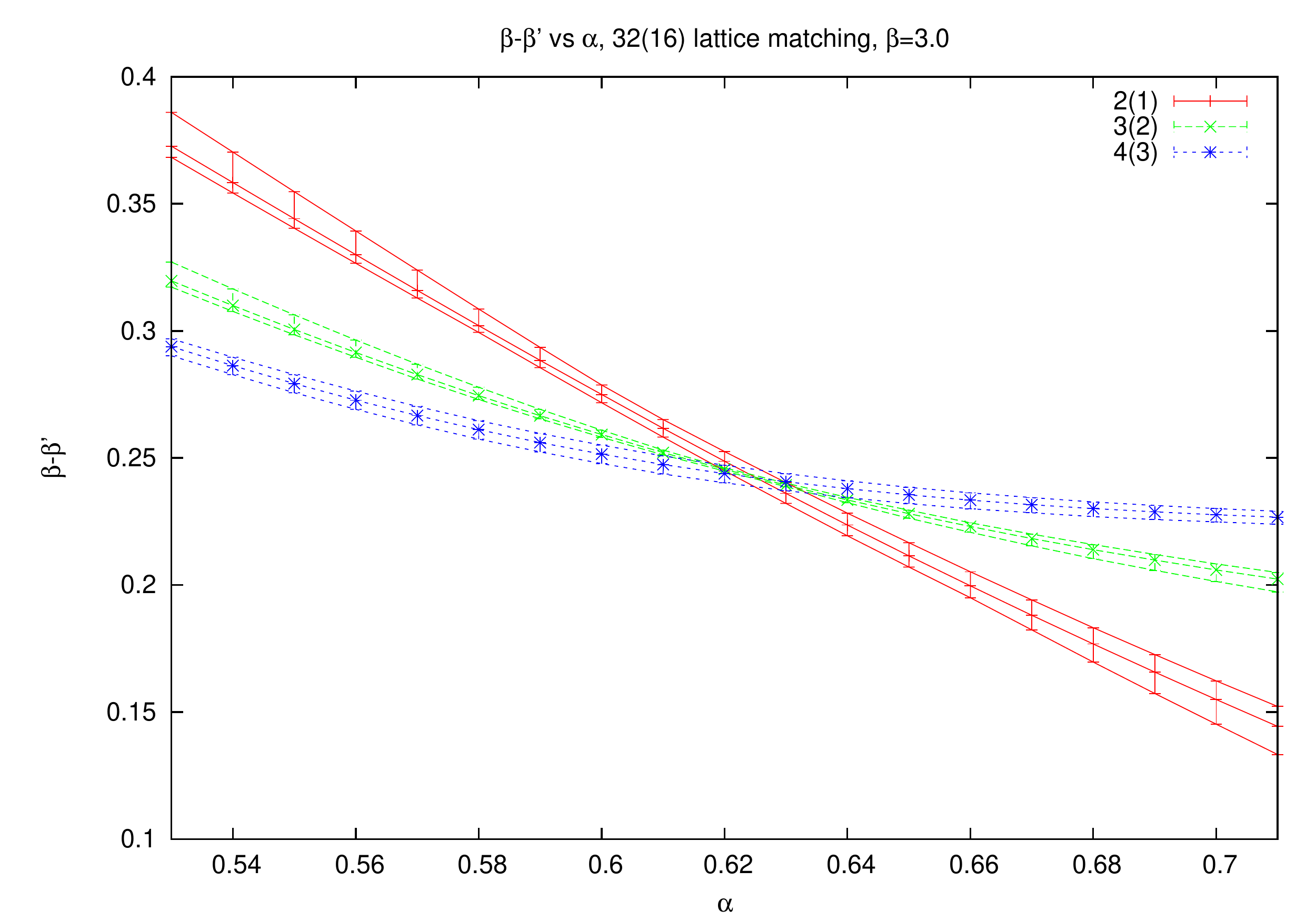}}
  \caption{An example of the matching of the plaquette in $\beta$ for the pure gauge case using ORIG blocking on 32(16) lattices. This is repeated for each observable to give a systematic error for each matching, then $\alpha$ is varied such that all blocking steps predict the same matching.}
  \label{fig:PURE_plaq}
\end{figure}

An example of the matching of the plaquette is shown in Fig.~\ref{fig:PURE_plaq_a}. The red, green and blue horizontal lines show the average plaquette on the $32^4$ lattice after 2, 3 and 4 ORIG blocking steps respectively, at $\beta=3.0, \alpha=0.57$. The interpolated red, green and blue points show the average plaquette on the $16^4$ lattice after 1, 2 and 3 blocking steps respectively, as a function of $\beta'$, also at $\alpha=0.57$. The value of $\beta'$ where the two red lines intersect gives the matching coupling for $n=2$, similarly the green and blue lines give the matching coupling for $n=3$ and $4$.

\begin{figure}[ht]
  \centering
\includegraphics[angle=0,width=14cm]{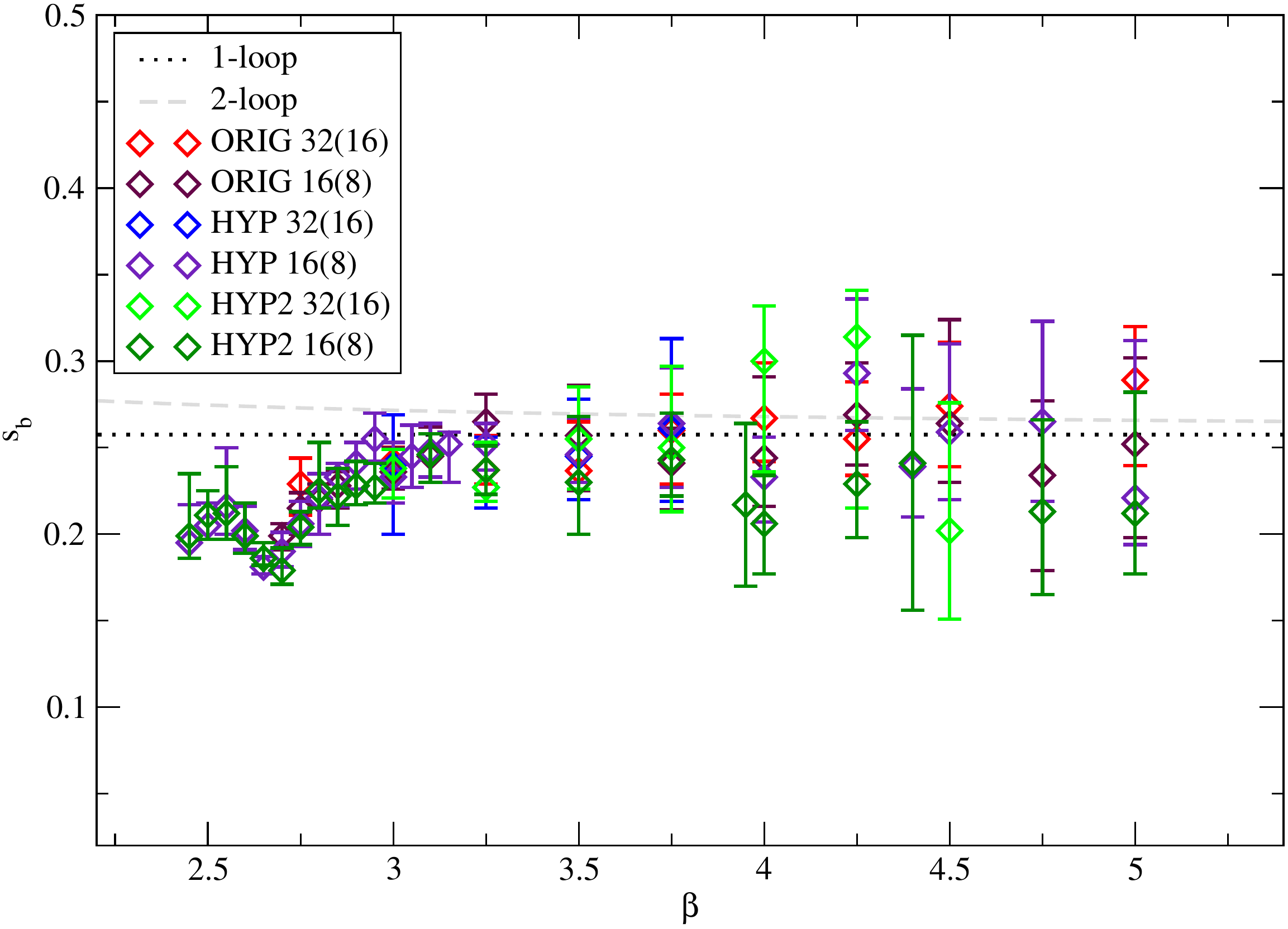}
  \caption{Bare step scaling $s_b$ for the pure gauge theory. Matching is performed on 32(16) and 16(8) lattices using ORIG, HYP and HYP2 blocking transforms. All give results consistent with each other and with perturbation theory.}
  \label{fig:PURE_sb}
\end{figure}

This matching is repeated for each observable, and the spread of predicted matchings for each $n$ gives a systematic error on the central matching value. The whole procedure is then repeated for various values of $\alpha$, as shown in Fig.~\ref{fig:PURE_plaq_b}, to find an optimal value of $\alpha$ where subsequent RG steps predict the same matching value. The intersection of the last two blocking steps gives a central value for $s_{b}(\beta=3.0;s=2)$, while the range of couplings for which any of the blocking steps intersect within errors gives the uncertainty on this central value. This was repeated for other values of $\beta$, and also with the HYP and HYP2 blocking transforms. Fig.~\ref{fig:PURE_sb} shows the resulting step scaling of the bare coupling $s_b$, determined using ORIG, HYP and HYP2 blocking on both 32(16) and 16(8) lattices, along with the 1--loop and 2--loop perturbative predictions. In the scaling region $s_b$ agrees well with the perturbative prediction, and the agreement between the different blocking transforms and lattice sizes shows that finite size errors are small, at least for the pure gauge case.

\FloatBarrier

\section{Anomalous Dimension Results}

Having confirmed that the two--lattice MCRG method works for the SU(2) pure gauge case, we now turn to the full Minimal Walking Technicolor theory of two Dirac fermions in the adjoint representation of SU(2).

There are two couplings of interest in this theory, the gauge coupling and the mass. At an IRFP the gauge coupling is expected to be irrelevant, leaving the mass as the only relevant operator. So we should in principle be able to match in the mass at arbitrary couplings, as long as we have sufficient RG steps for the gauge coupling to flow to its FP value, and the mass is small enough that despite the resulting flow in the mass the system remains close to the IRFP. In practice we only have a small number of RG steps, and because the beta function is small the coupling only flows slowly towards its FP value, so we set $\beta'=\beta$ assuming the flow in the coupling due to one RG step is negligible. The case $\beta'\neq\beta$ will be considered later. We match observables as for the pure gauge case, but instead of matching in $\beta$, we fix $\beta'=\beta$, and match pairs of bare masses $(am_0,a'm_0')$.

\begin{figure}[ht]
  \centering
\subfigure[Plaquette Matching]{\includegraphics[angle=270,width=7.5cm]{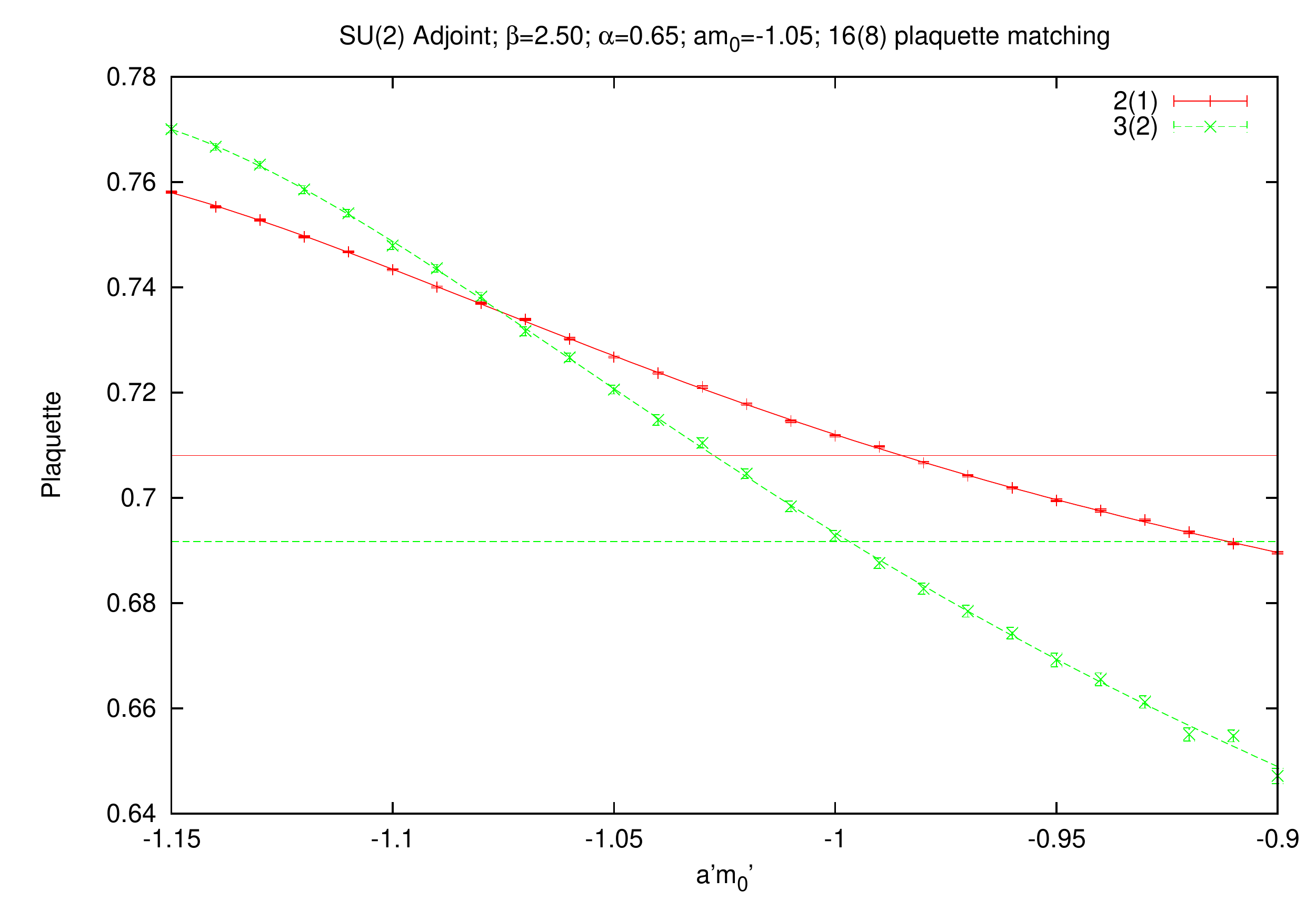}}
\subfigure[$\alpha$--Optimisation]{\includegraphics[angle=270,width=7.5cm]{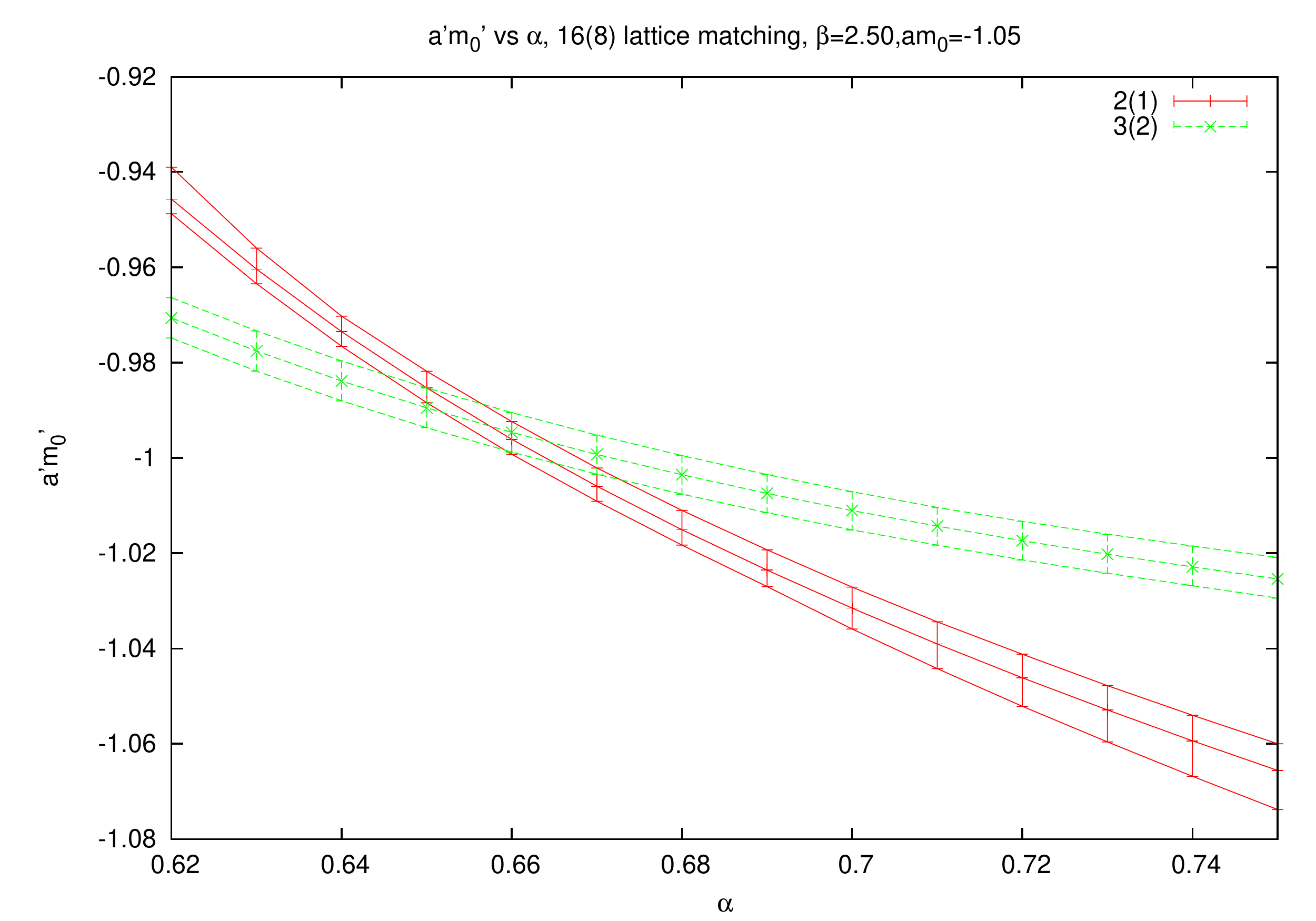}}
  \caption{An example of the matching of the plaquette in the bare mass $am_0$. This is repeated for each observable to give a systematic error for each matching, then $\alpha$ is varied such that all blocking steps predict the same matching.}
  \label{fig:MASS_plaq}
\end{figure}

We use the HiRep~\cite{DelDebbio:2008zf} implementation of the Wilson plaquette gauge action with adjoint Wilson fermions and an RHMC algorithm with two pseudofermions. We generated $\sim3000$ configurations on $16^4$ and $8^4$ lattices, for a range of bare masses at each $\beta$. This allows two matching steps, after $2(1)$ and $3(2)$ steps on the $16^4(8^4)$ lattices respectively. See Fig.~\ref{fig:MASS_plaq} for an example of this matching and subsequent $\alpha$--optimisation.

\begin{figure}[ht]
  \centering
  \includegraphics[angle=270,width=10cm]{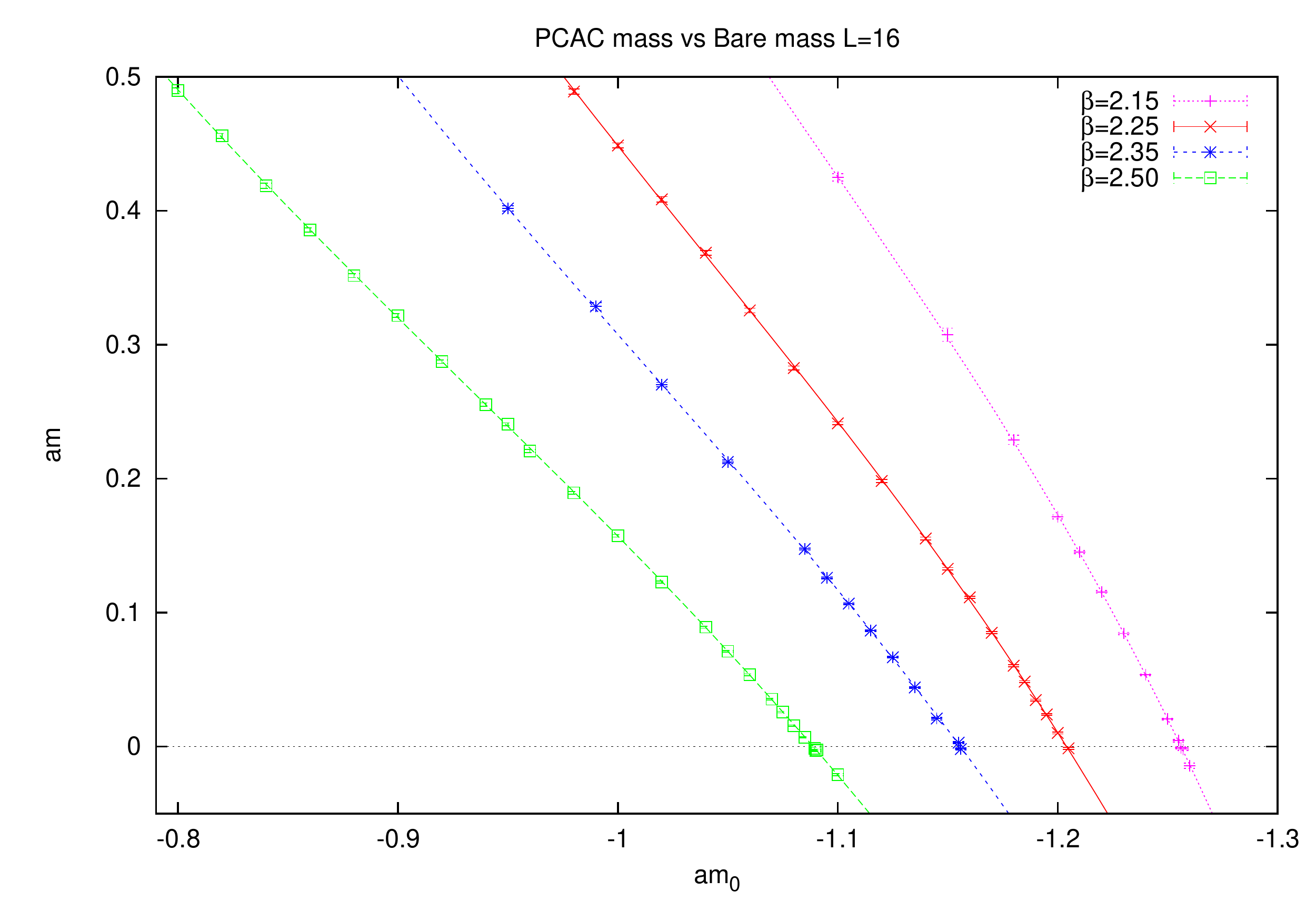}
  \caption{PCAC mass, $am$, as a function of the bare mass, $am_0$, on $16^4$ lattices for $\beta=2.15,2.25,2.35,2.50$}
  \label{fig:MASS_pcac}
\end{figure}

\begin{figure}[ht]
  \centering
\includegraphics[angle=270,width=12.0cm]{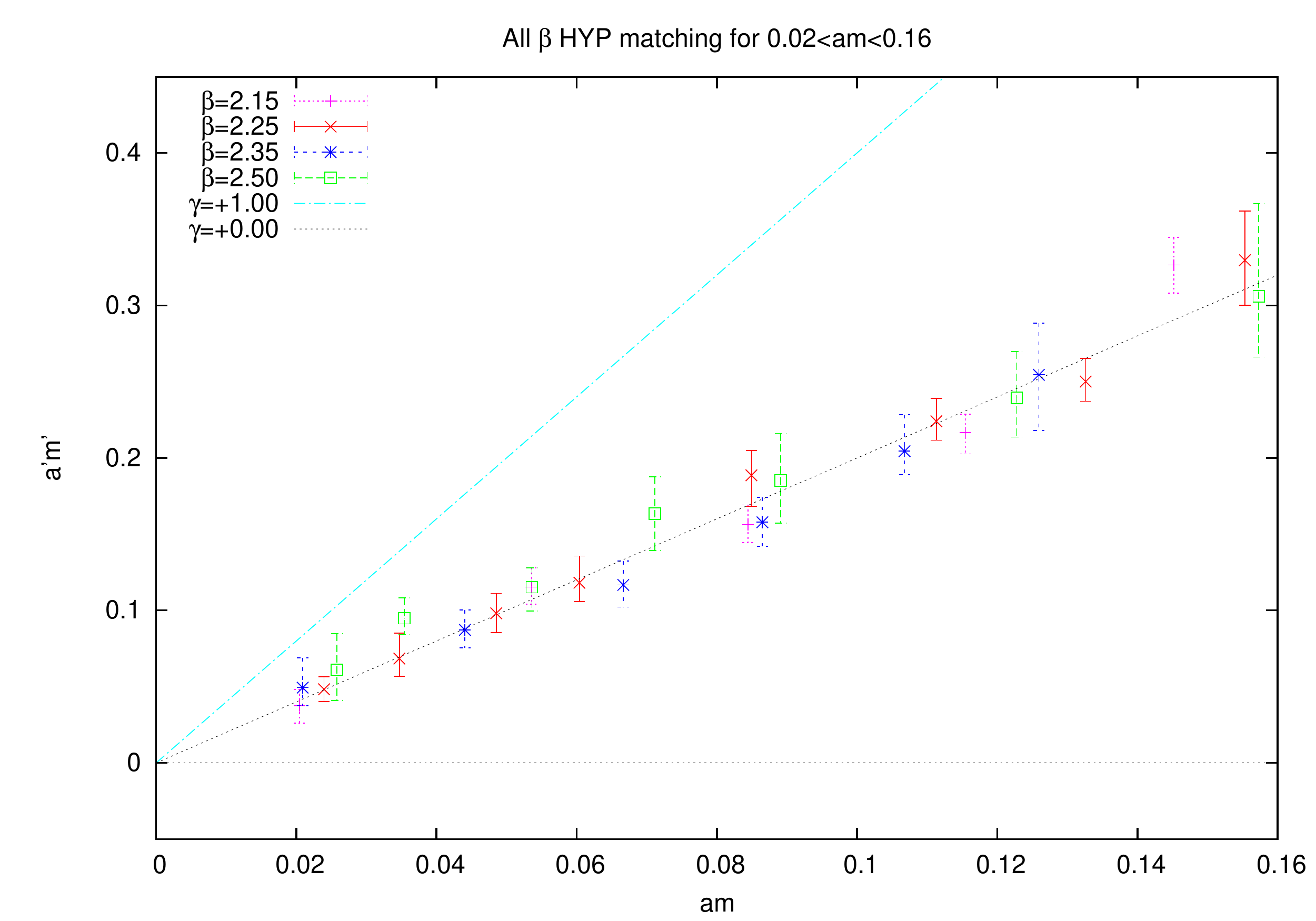}
  \caption{HYP Matching in mass using all $\beta$ values in the mass range $0.02<am<0.16$. Consistent with a vanishing anomalous mass dimension, $\gamma=1$ is strongly disfavoured.}
  \label{fig:MASS_HYPall}
\end{figure}

Because the bare mass is additively renormalised we convert the bare masses to PCAC masses,
\begin{equation}
am \equiv a m_{\mathrm{PCAC}}\left(\tfrac{L}{2}\right), \quad a m_{\mathrm{PCAC}}(x_0) = \frac{aD^{s}_0f_{A}(x_0)}{2f_{P}(x_0)}.
\end{equation}
where $D^{s}_0$ is the symmetric lattice derivative operator, and $f_A$ and $f_P$ are axial and pseudoscalar correlators respectively.
We measure the PCAC mass, $am$, as a function of bare mass, $am_0$, for each $\beta$ on the $16^4$ lattices, as shown in Fig.~\ref{fig:MASS_pcac}. We then use this to convert the bare masses on both $8^4$ and $16^4$ lattices to PCAC masses, as the measured PCAC masses on the $8^4$ lattices suffer from finite volume effects, as described in App.~\ref{app:pcac}. Our previous result~\cite{Catterall:2010du} for the anomalous mass dimension used PCAC masses measured on the $8^4$ lattices and hence contained a large finite volume effect, which has been removed in the present work.
The anomalous mass dimension appears in the RG equation for the mass
\begin{equation}
\frac{d (am)}{d\ln|\mu|} = -y_m am = -(1+\gamma)am.
\end{equation}
At an IRFP the anomalous mass dimension is a constant, so the expression can be integrated to give
\begin{equation}
\label{eq:gamma}
\frac{a'm'}{am} = 2^{\gamma+1}
\end{equation}
for a pair of matching masses $(am,a'm')$, from which a value for $\gamma$ can be extracted. We used four values of $\beta$, $\beta=2.15,2.25,2.35,2.50$, and the matching PCAC mass pairs using the HYP blocking tranform are shown in Fig.~\ref{fig:MASS_HYPall}. We also repeated the matching using ORIG and HYP2 blocking. A comparison of the three blocking methods is shown for each $\beta$ value in Fig.~\ref{fig:MASS_blocking}, and the agreement between them is very good.

Different $\beta$ values seem to predict consistent values for the anomalous mass dimension, as shown in Fig.~\ref{fig:MASS_HYPall}, which uses all the beta values and masses in the range $0.02<am<0.16$. Masses above this range are excluded because the data are no longer consistent with a linear fit, indicating that we are beyond the small--mass linear scaling region. The masses below $am=0.02$ are excluded because they are likely to contain a significant unwanted contribution from the running of the coupling, as described in App.~\ref{app:mass}.

\begin{figure}[ht]
  \centering
\includegraphics[angle=270,width=12.0cm]{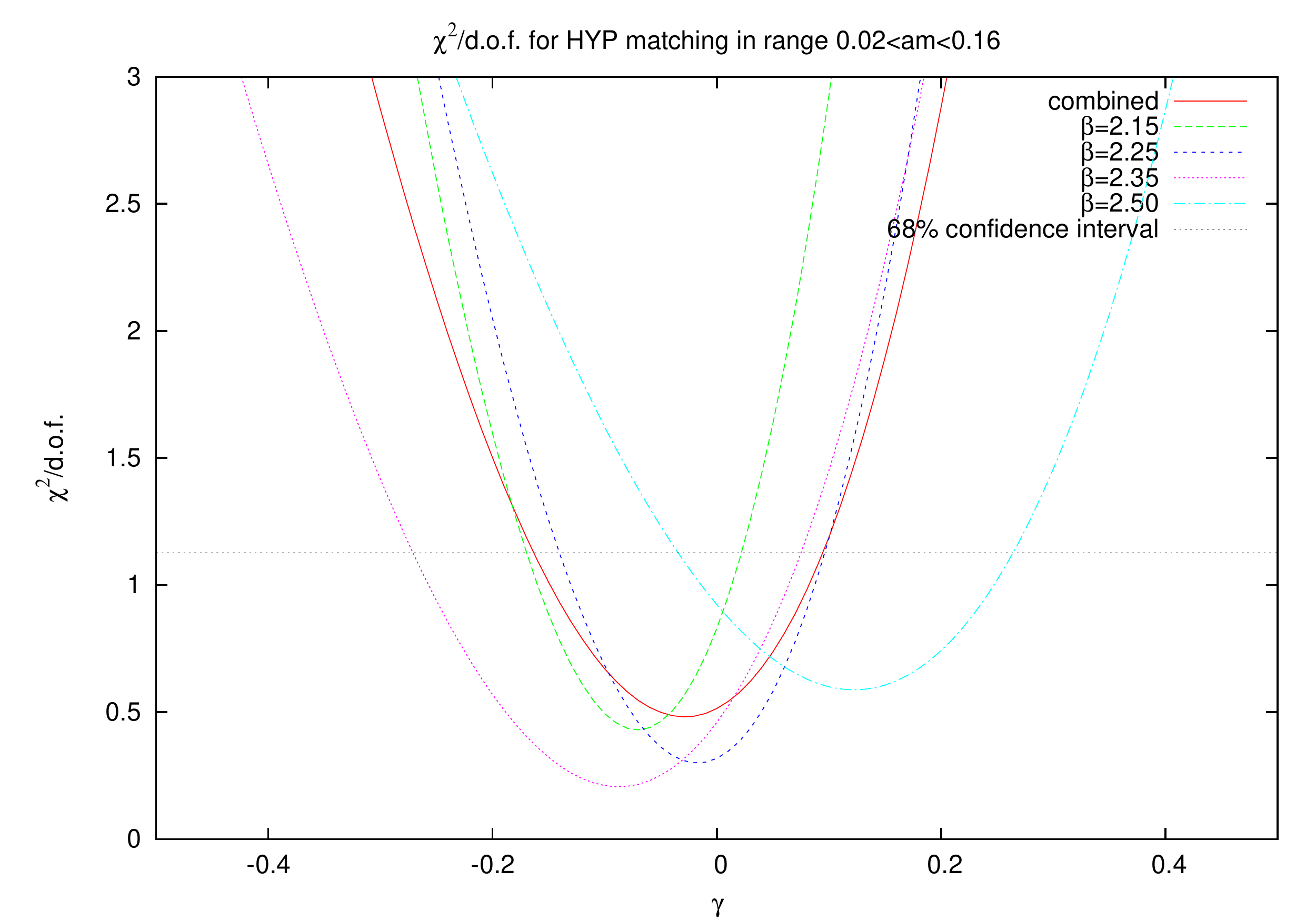}
  \caption{$\chi^2/\mathrm{d.o.f.}$ of $\gamma$ with HYP Matching in mass using all $\beta$ values in the mass range $0.02<am<0.16$. Combined best fit with $22$ $\mathrm{d.o.f.}$ gives $\gamma=-0.03(13)$.}
  \label{fig:MASS_HYPchisq}
\end{figure}

\begin{figure}[ht]
  \centering
\includegraphics[angle=270,width=6.0cm]{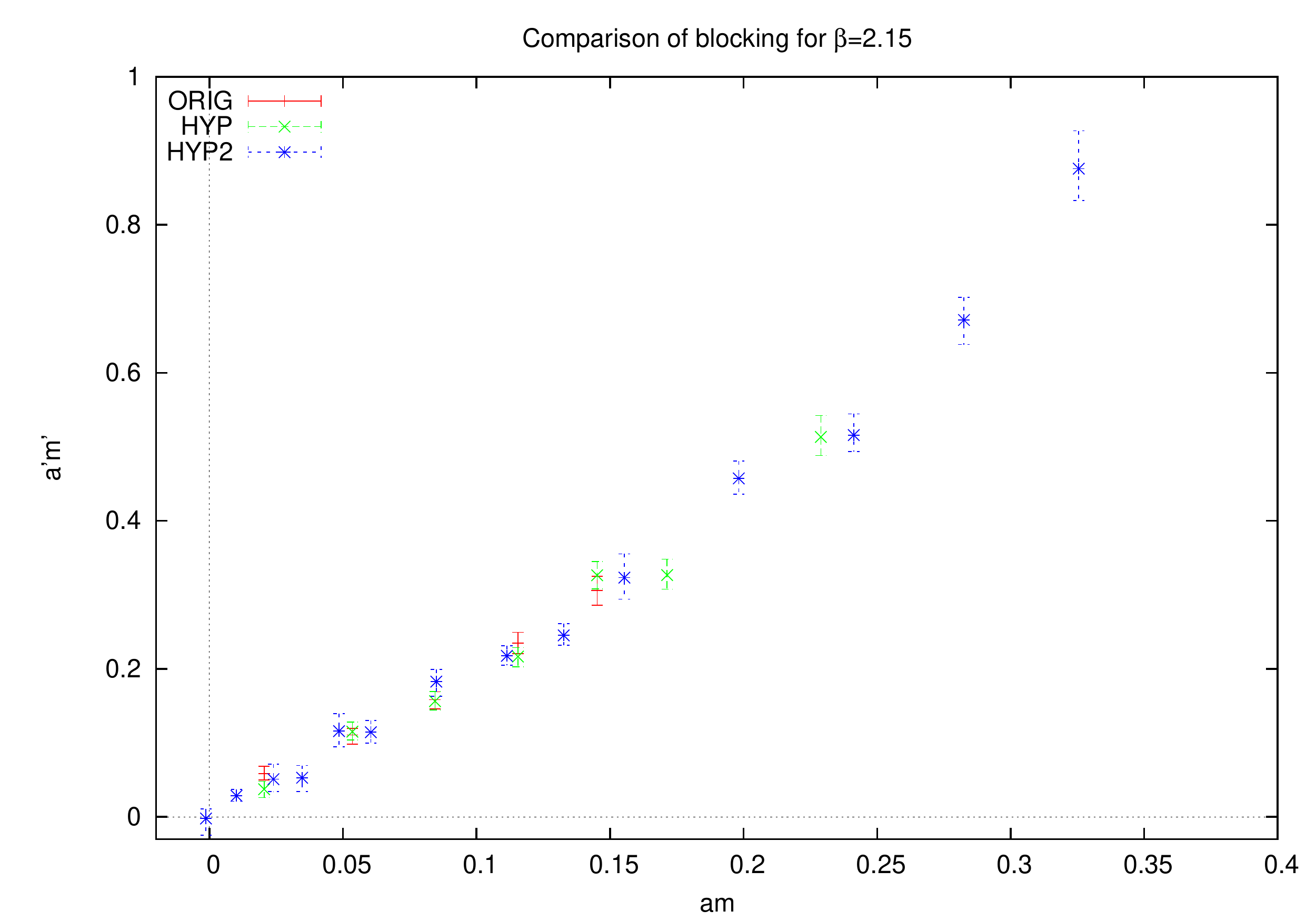}
\includegraphics[angle=270,width=6.0cm]{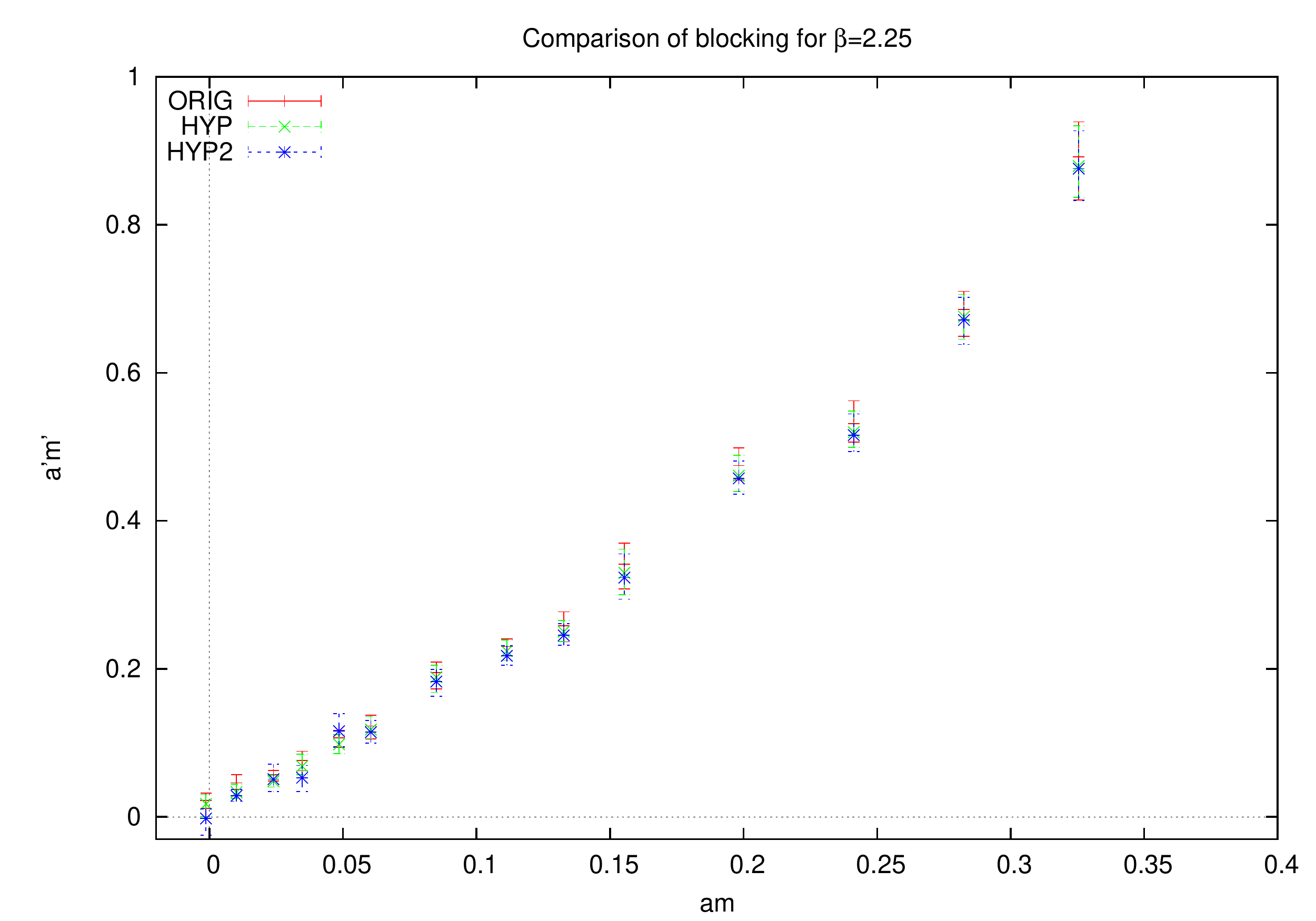}

\includegraphics[angle=270,width=6.0cm]{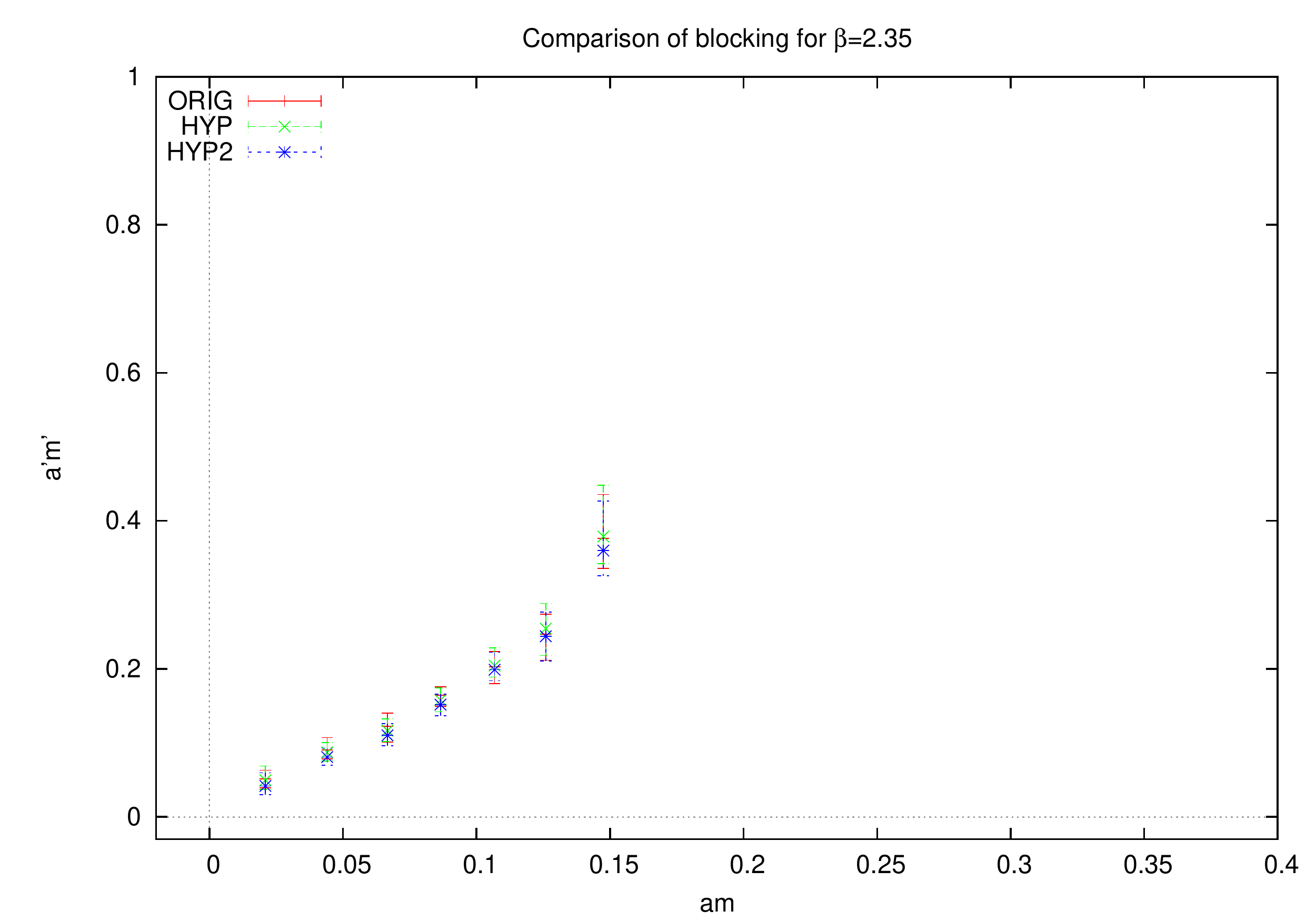}
\includegraphics[angle=270,width=6.0cm]{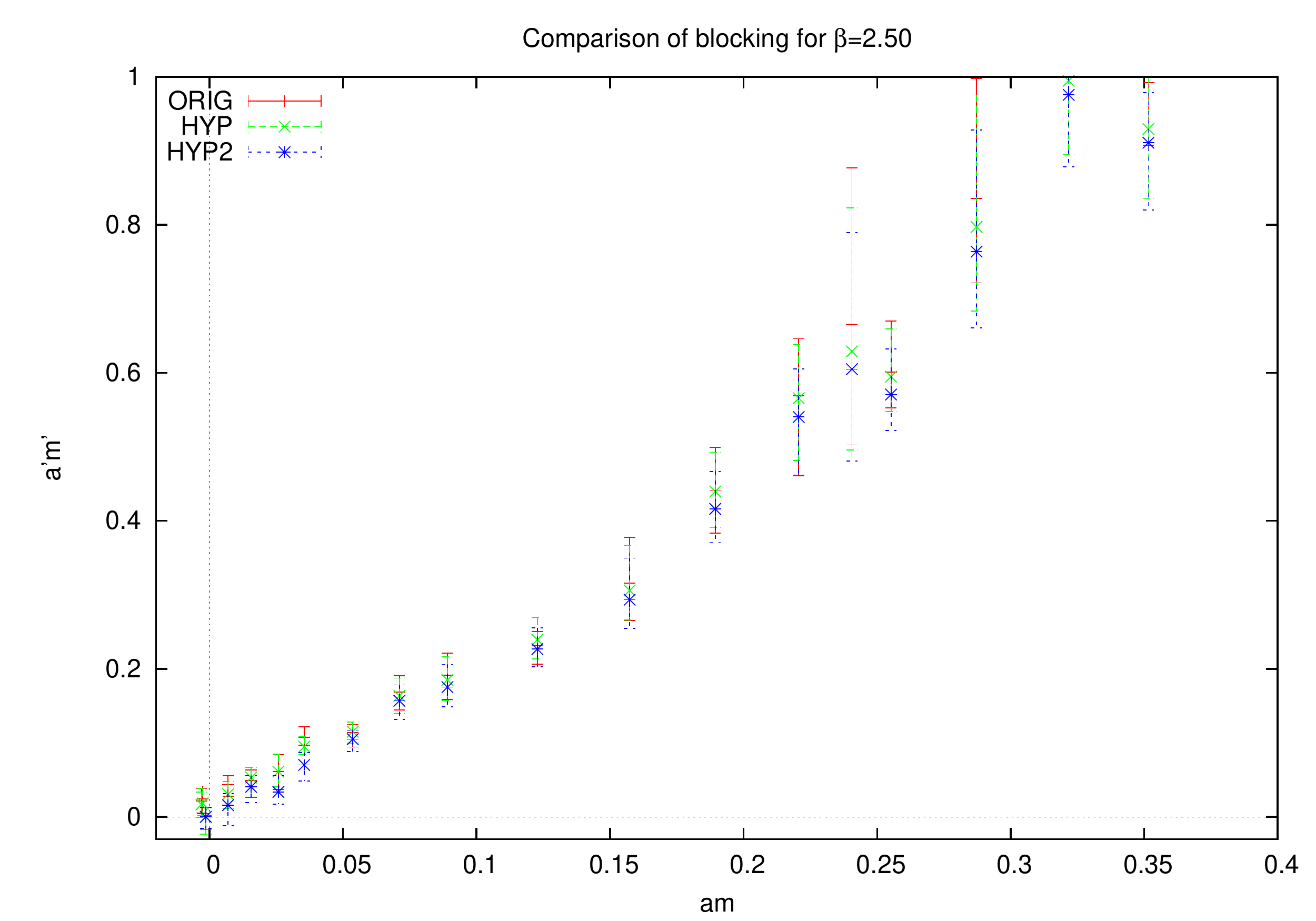}
  \caption{Comparison of matching mass pairs using different blocking transforms.}
  \label{fig:MASS_blocking}
\end{figure}

The $\chi^2/\mathrm{d.o.f.}$ for a linear fit of the form in Eq.~\ref{eq:gamma} is shown in Fig.~\ref{fig:MASS_HYPchisq}, which favours a vanishing anomalous mass dimension. A combined fit to all $\beta$ and $m$ gives $\gamma=-0.03(13)$, with $\gamma>0.13$ ruled out at a 95\% confidence level. Significant systematic errors exist that have not been accounted for in this result, in particular the assumption that we can set $\beta'=\beta$. These systematic errors will be discussed in Sec.~\ref{sec:systematics}.

\[\begin{array}{c|c|c}
\beta & \gamma & \mathrm{d.o.f.} \\
\hline
2.15 & -0.07(10) & 3 \\
2.25 & -0.02(12) & 6 \\
2.35 & -0.09(17)& 5 \\
2.50 & 0.12(15)& 5 \\
\hline
\mathrm{all} & -0.03(13) & 22 \\
\end{array}\]

\FloatBarrier

\section{Discrete Beta Function Results}
\label{sec:coupling}

It would also be interesting to directly measure the step scaling of the bare coupling $s_b(\beta,s)$ for the full theory, as was done for the pure gauge case. A fixed point would be indicated by a change of sign in this quantity as the bare coupling is varied from weak to strong coupling. The difficulty is that the mass is a relevant operator, while the coupling is expected to be at best nearly marginal, so in order for the MCRG to pick out the behaviour of the coupling the mass would have to be tuned to zero.
Furthermore, even if the mass is tuned sufficiently close to zero that we are initially following the evolution of the gauge coupling (assuming it is the least irrelevant remaining operator), we can no longer take $n\rightarrow \infty$ limit, because the coupling will flow to its FP value, and the flow of the mass will eventually dominate.

So the method is to tune the mass close to zero initially, then take a few RG steps, where the flow is hopefully following the gauge coupling, and extract the running of the coupling from this, before the flow in the mass becomes significant.

\begin{figure}[ht]
  \centering
\includegraphics[angle=270,width=12.0cm]{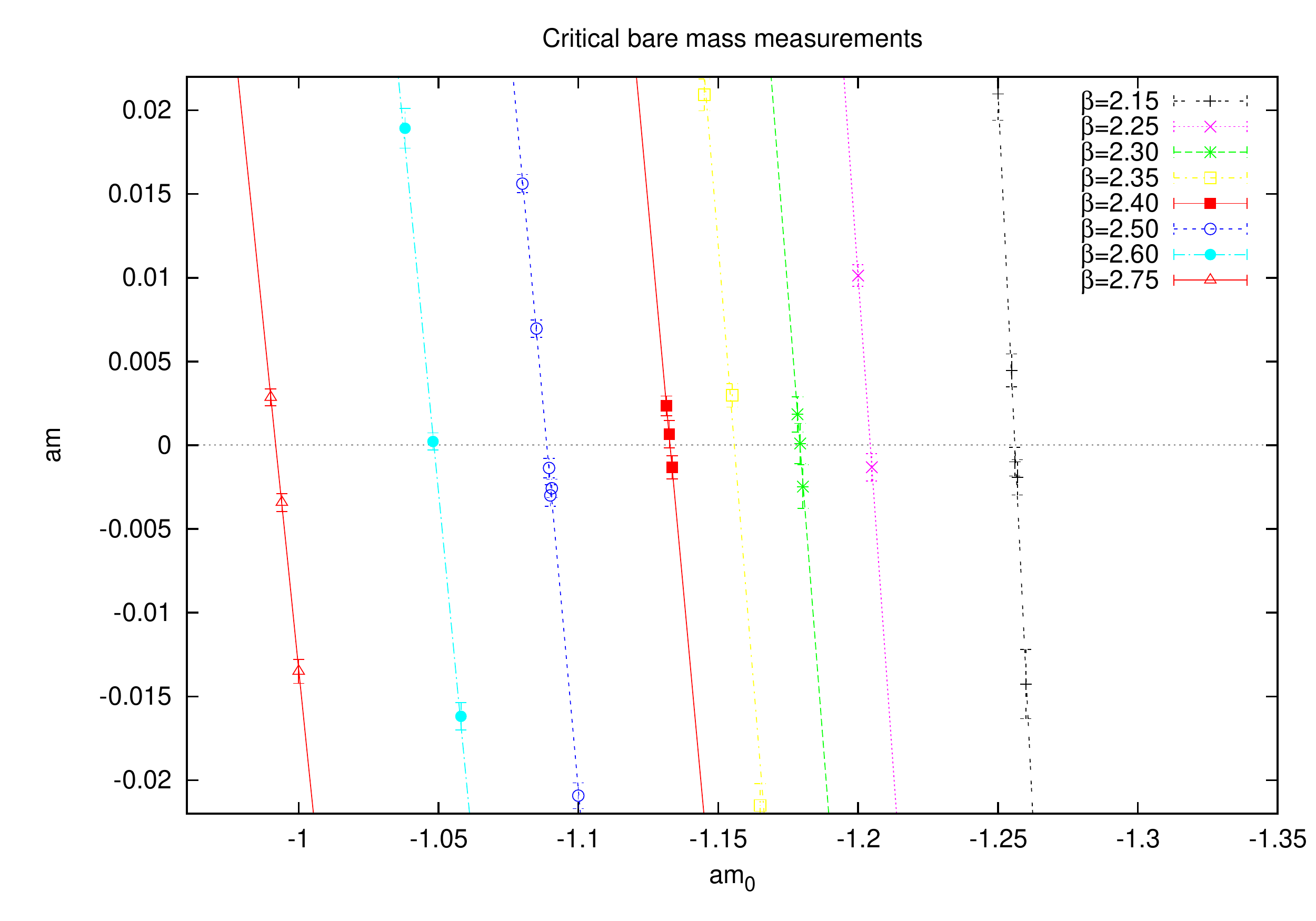}
  \caption{Measurement of critical $m_0$ values from $16^4$ lattices.}
  \label{fig:MASSLESS_measurement}
\end{figure}

We measure the PCAC mass on $16^4$ lattices for a range of $\beta$ values at small masses, and for each $\beta$ extract the critical bare mass using a linear interpolation in the PCAC masses, as shown in Fig.~\ref{fig:MASSLESS_measurement}.

\begin{figure}[ht]
  \centering
\includegraphics[angle=270,width=12.0cm]{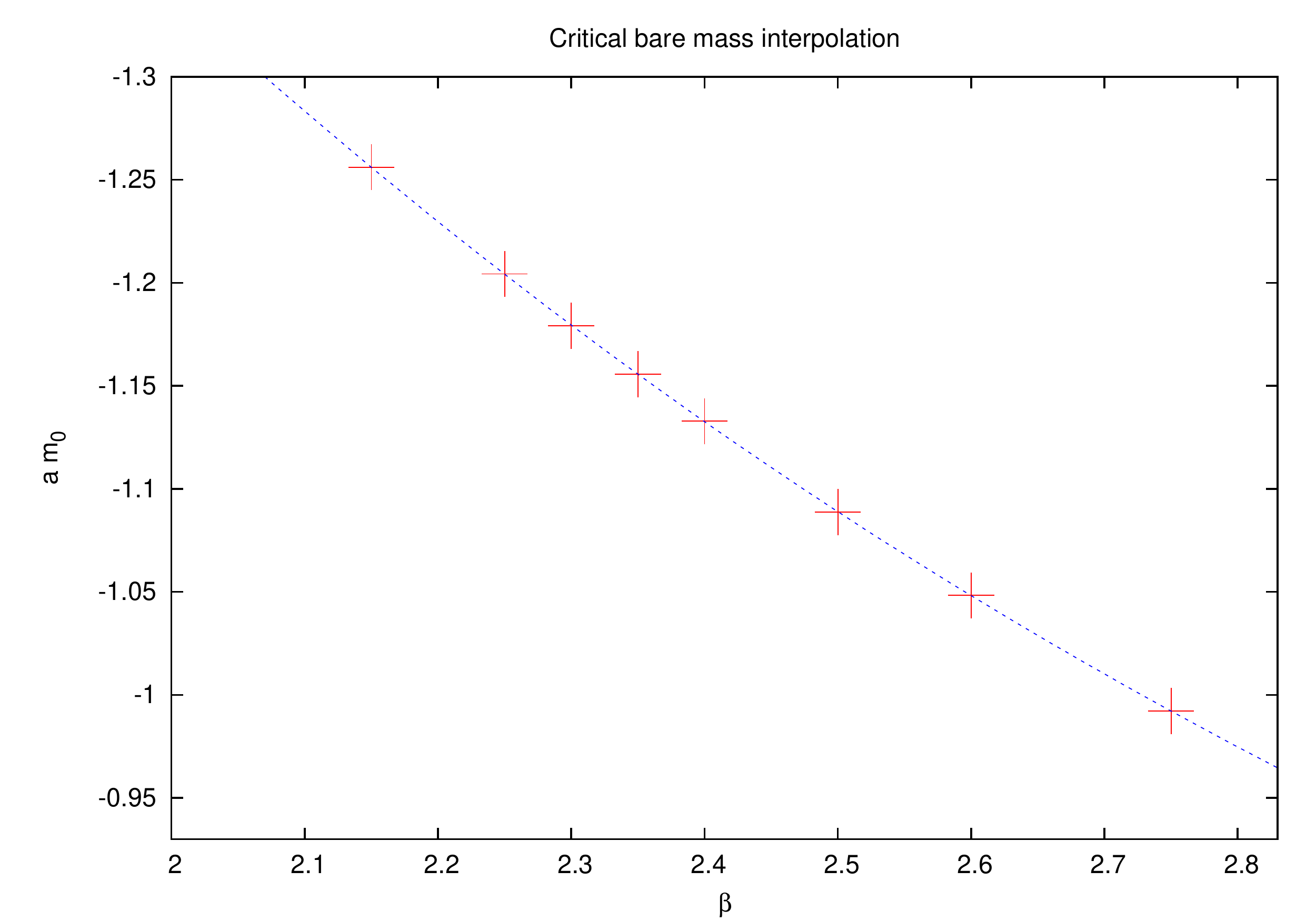}
  \caption{Interpolation of critical $am_0$ values from $16^4$ lattices, to be used to determine critical $am_0$ values for the massless $8^4$ runs.}
  \label{fig:MASSLESS_interpolation}
\end{figure}

We then fit an interpolating function in $\beta$ to these critical masses, as shown in Fig.~\ref{fig:MASSLESS_interpolation}, which gives us the critical bare mass for any $\beta$ in the range $2.15< \beta < 2.75$. The systematic error in the PCAC mass due to this interpolation is small, $\lesssim 0.001$, as described in App.~\ref{app:massless}.

\begin{figure}[ht]
  \centering
\includegraphics[angle=270,width=12.0cm]{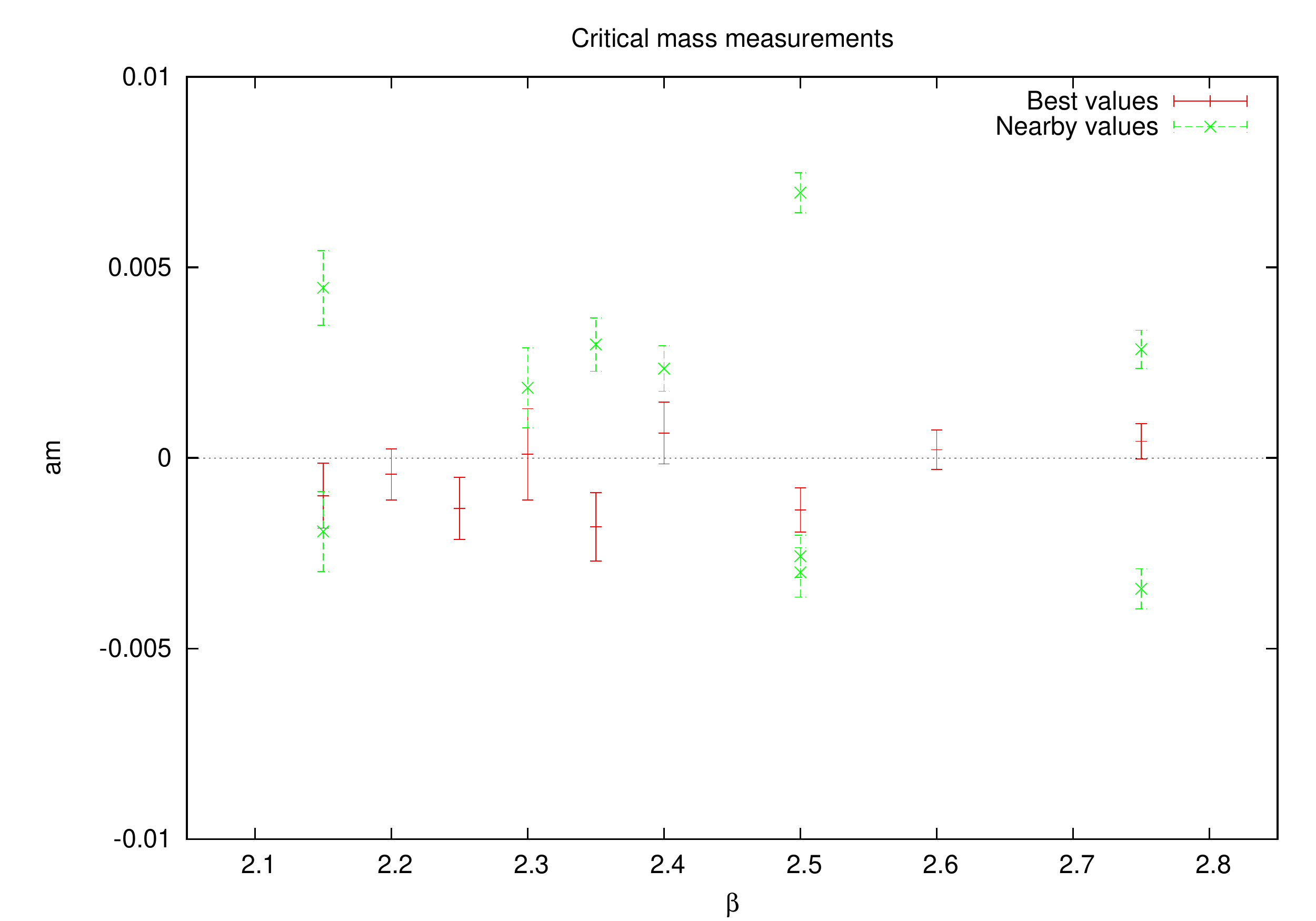}
  \caption{Measured PCAC mass of critical $16^4$ runs used for massless matching.}
  \label{fig:MASSLESS_bestmeasurement}
\end{figure}

The measured PCAC masses for the $16^4$ critical runs are shown in Fig.~\ref{fig:MASSLESS_bestmeasurement}, along with some runs at small masses to allow us to check for any systematic mass--dependence.

\begin{figure}[ht]
   \centering
 \subfigure[Plaquette Matching]{\includegraphics[angle=270,width=7.5cm]{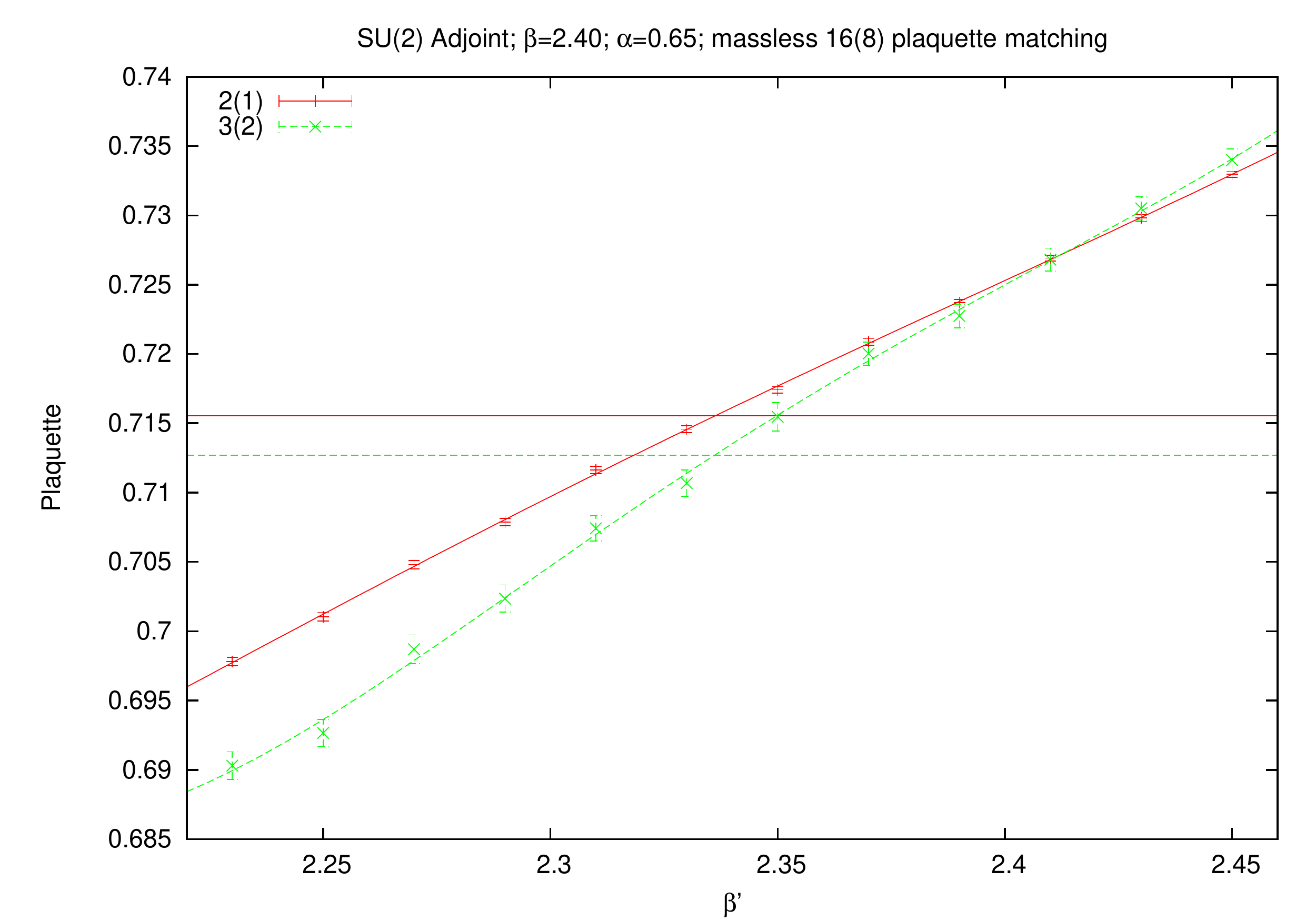}}
 \subfigure[$\alpha$--Optimisation]{\includegraphics[angle=270,width=7.5cm]{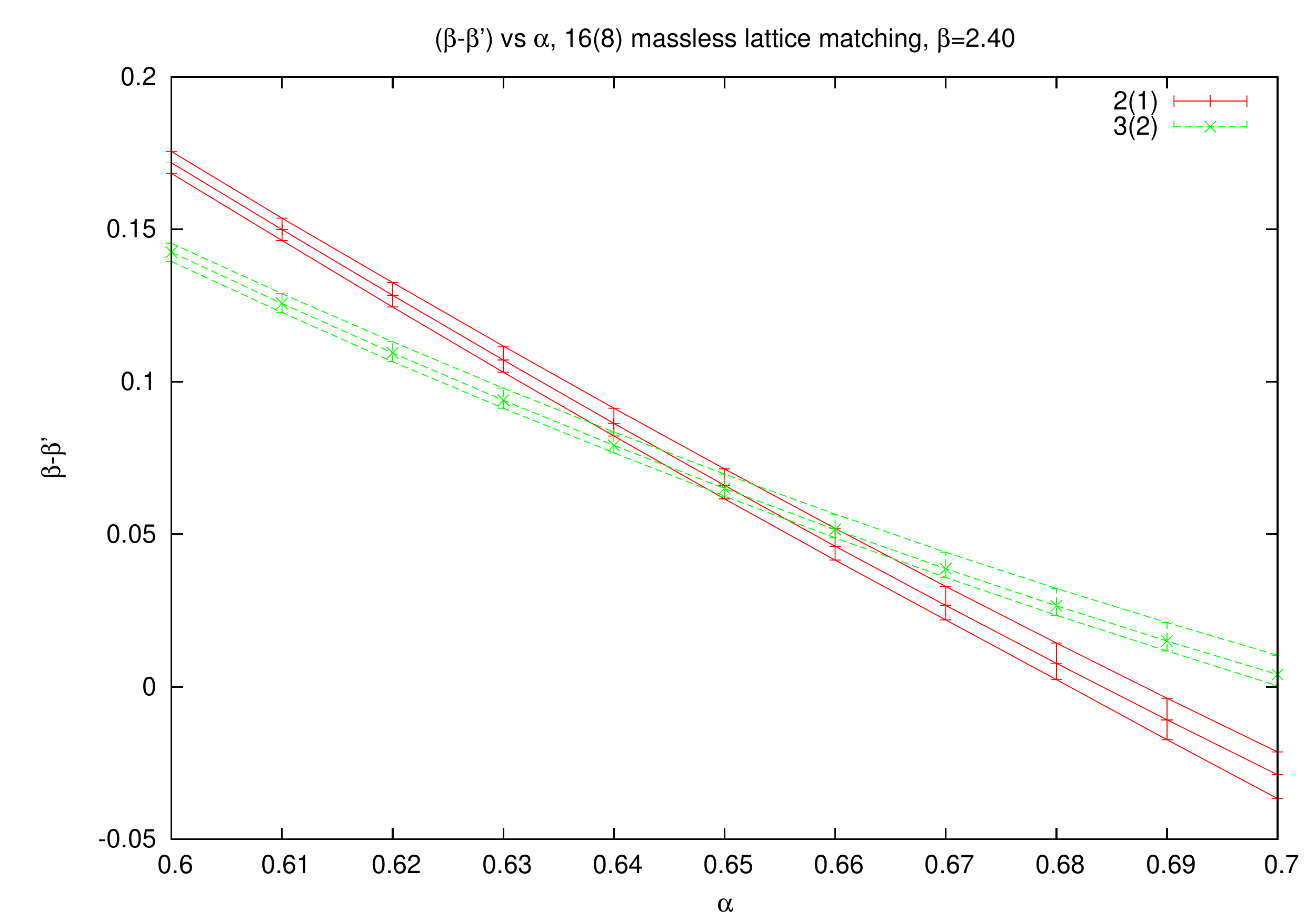}}
\caption{An example of the matching of the plaquette in $\beta'$ for the massless case using ORIG blocking. This is repeated for each observable to give a systematic error for each matching, then $\alpha$ is varied such that all blocking steps predict the same matching.}
   \label{fig:MASSLESS_plaq}
 \end{figure}

We generated $\sim3000$ configurations on $16^4$ and $8^4$ lattices for a range of $\beta$ values, each run at the critical bare mass. The matching procedure in $\beta$ is then essentially the same as for the pure gauge case. An example of the matching in the plaquette, and subsequent $\alpha$--optimisation is shown in Fig.~\ref{fig:MASSLESS_plaq}.

\begin{figure}[ht]
   \centering
 \includegraphics[angle=270,width=13.0cm]{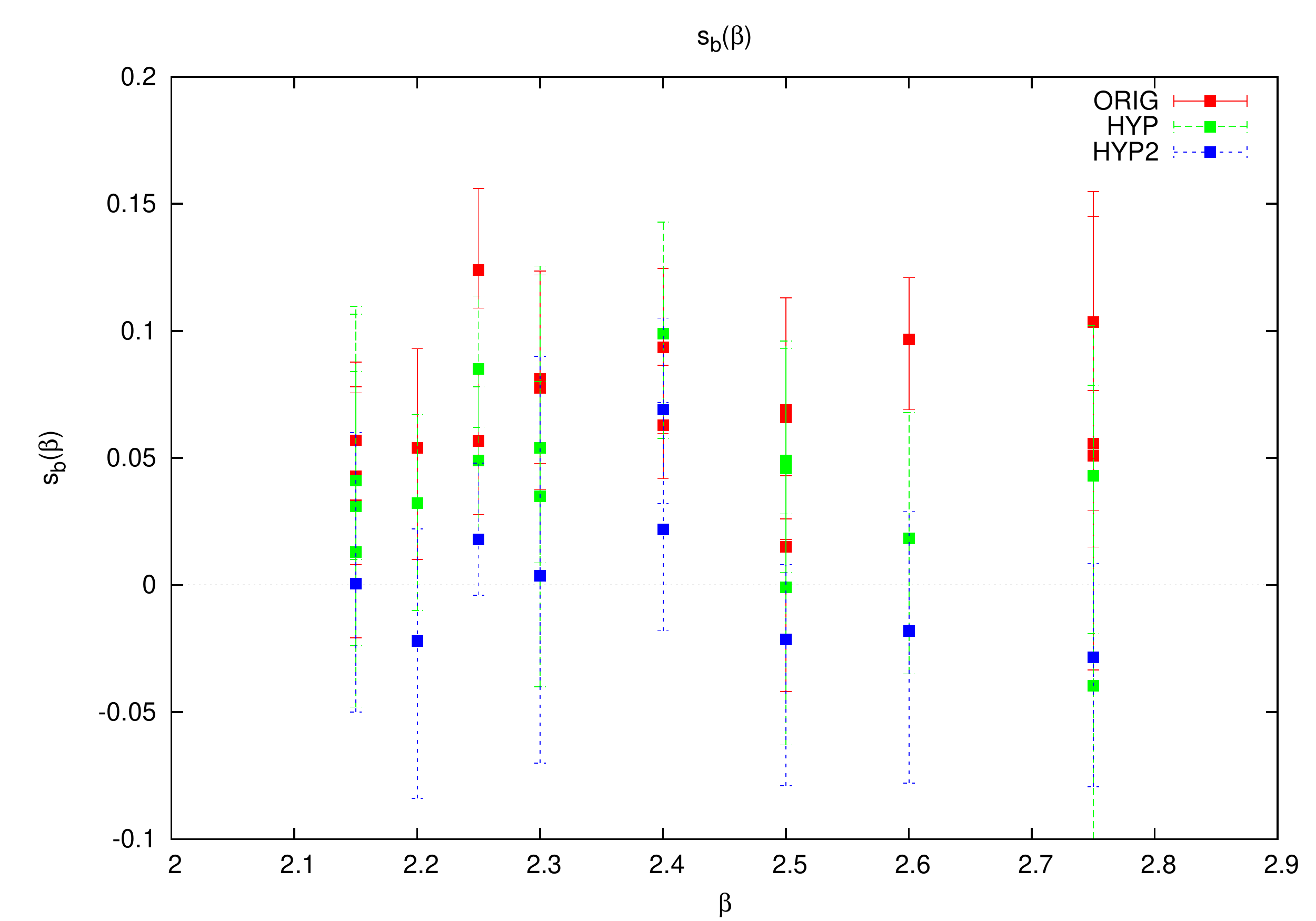}
   \caption{Massless matching in $\beta$ on 16(8) lattices. This includes both the massless and some small mass runs at each $\beta$ - within errors $s_b$ shows no mass dependence for these small masses.}
   \label{fig:MASSLESS_sb}
 \end{figure}

The resulting measurement of $s_b(\beta)$ is shown in Fig.~\ref{fig:MASSLESS_sb}. This includes both the massless and small mass $16^4$ runs - within errors $s_b$ shows no mass dependence for these small masses. The ORIG matching values of $s_b$ are clearly positive throughout, the HYP values are lower, and the HYP2 values are consistent with zero within error bars. There is no clear cross--over from positive to negative values of $s_b$ for any of the blocking transforms, so while the data are consistent with a fixed point, they are not sufficiently precise to distinguish slow running from a fixed point. This level of precision is nonetheless similar to that found in the Schrodinger Functional studies, albeit with less understood systematic errors.

\FloatBarrier

\section{Systematic Errors}
\label{sec:systematics}

\subsection{Matching Observables}

In principle, for a given $(\beta,am)$ there should be a unique matching set of couplings $(\beta',a'm')$, for which all blocked observables agree after $n$ and $(n-1)$ blocking steps respectively. In this work we have set $\beta=\beta'$ and we were able to find $a'm'$ such that the blocked observables matched.

\begin{figure}[ht]
  \centering
\includegraphics[angle=270,width=12.0cm]{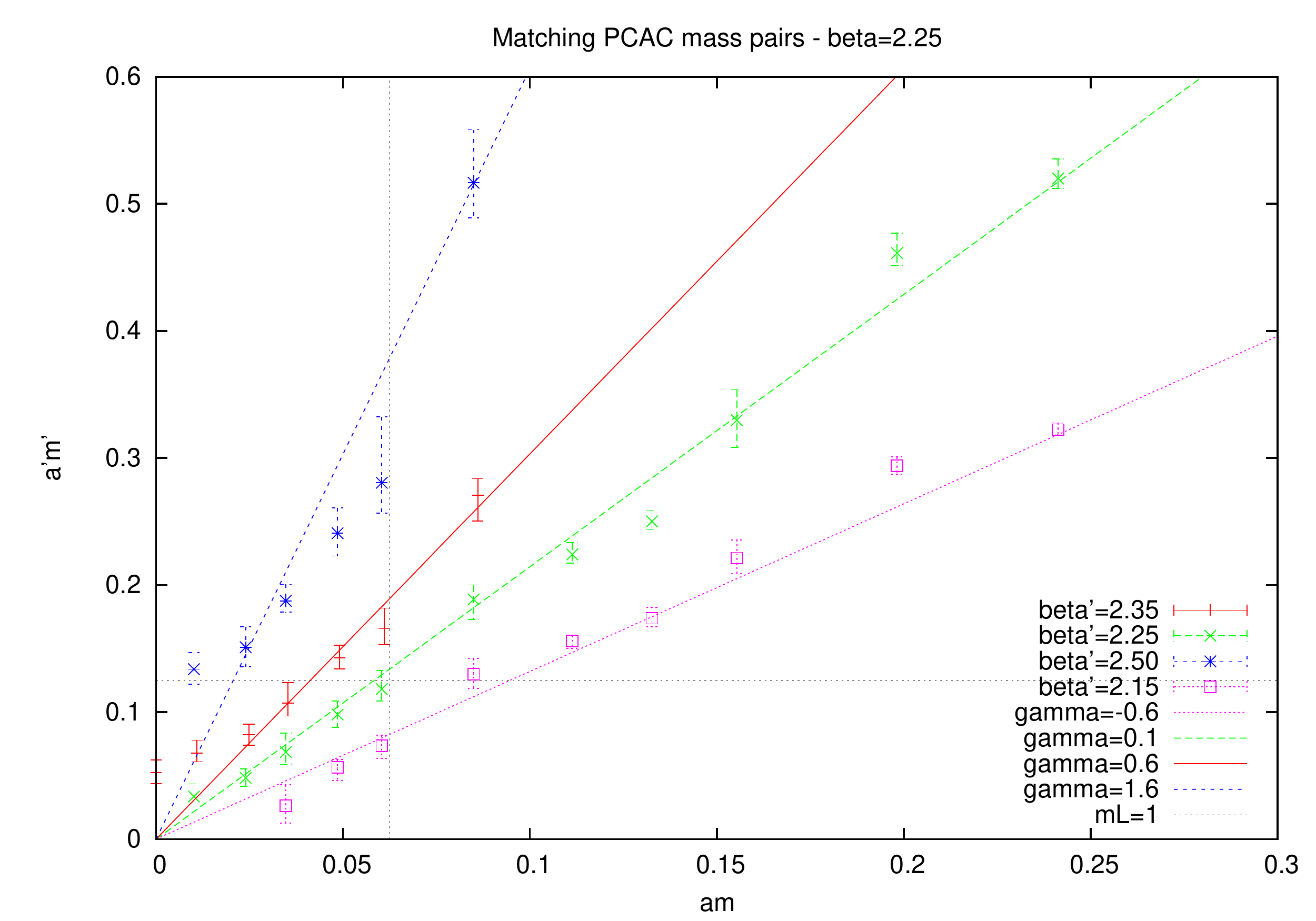}  
\caption{Mass matching pairs at $\beta=2.25$ for a range of $\beta'$. Varying $\beta'$ can lead to very different results for $\gamma$. The dotted horizontal and vertical lines show $mL,m'L'=1$.}
  \label{fig1}
\end{figure}

In practice however, all of our observables are small Wilson loops, and as such are strongly correlated and have a very similar dependence on $\beta'$ and $a'm'$. This means that we can in fact find a ``matching'' $a'm'$ for a range of values of $\beta'$, which, given that we do not know the correct value of $\beta'$ to use, significantly increases the error on our determination of $\gamma$. As an example the matching mass pairs for $\beta=2.25$ and various values of $\beta'$ are shown in Fig.~\ref{fig1}.

In Sec.~\ref{sec:coupling}, while we find that $s_b=\beta-\beta'$ is compatible with zero, corresponding to setting $\beta=\beta'$, the error bars are relatively large, enclosing the region $-0.08\lesssim \beta-\beta' \lesssim 0.16$. From Fig.~\ref{fig1} we see that for $\beta=2.25$ this region is approximately bounded by $\beta'=2.15$ and $\beta'=2.35$, and encloses a large range of values for the anomalous mass dimension, $-0.6 \lesssim \gamma \lesssim 0.6$. This range is representative of the errors in the anomalous mass dimension due to the uncertainty in the correct value of $\beta'$, and is the dominant source of systematic uncertainty in our results.

To resolve this issue, more observables which are ``orthogonal'' to the Wilson loops, for example meson correlators or other fermionic observables such as the PCAC mass, would need to be included in the matching determination to determine which of these blocked configurations are actually matched and find a unique matching set of couplings. We were unable to consider these observables with our current lattice volumes due to the very small size ($2^4$ and $4^4$) of the blocked configurations on which the observables are measured.

\subsection{Finite--volume Effects}
The MCRG techniques that have been described previously have achieved much success
when applied to QCD--like theories. However, previous studies both analytical and
numerical of the minimal walking model studied in this paper suggest that it is likely
conformal or very nearly conformal in the infrared.
Thus the calculations of $\gamma$ require us to compute the flow in the 
mass parameter for a conformal field theory (CFT) in the presence of a small mass deformation.

\begin{figure}[ht]
  \centering
\includegraphics[angle=270,width=12.0cm]{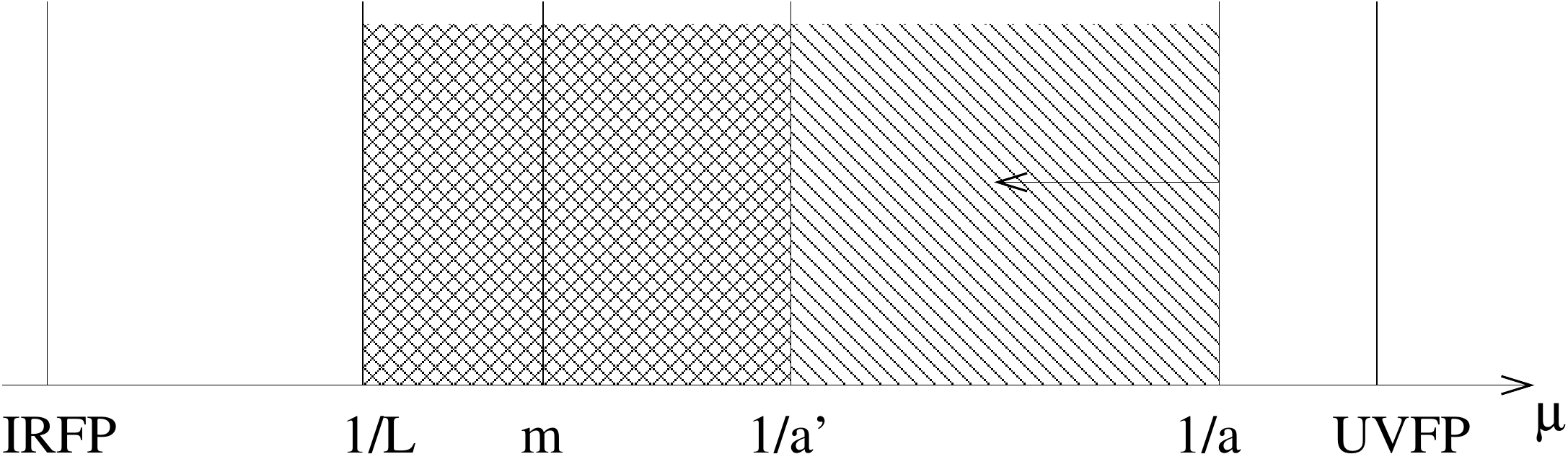}  
\caption{Separation of scales in a normal lattice simulation. The lattice spacing $a$ provides a UV--cutoff $1/a$, the lattice size $L$ provides an IR--cutoff $1/L$, and the mass of the lightest propagating state $m$ lies somewhere between these two scales. After an RG blocking step the lattice spacing changes to $a'=2a$, and the modes between $1/a'$ and $1/a$ are integrated out.}
  \label{fig:scale1}
\end{figure}

Of course to be a true CFT the theory must be considered in infinite volume but
unfortunately when we do simulations we have to live with
lattices of finite size and this makes the analysis a little more subtle~\cite{DelDebbio:2010ze}.
In principle both the mass parameter and the lattice size then
determine any correlation length.
If we want to extract the correct physics for the
infinite volume CFT it is important to make sure that
the correlation lengths we are measuring and matching are not being
strongly influenced by the finite box size.
So we should make sure that the physical lattice size
is much larger than the correlation length or equivalently $m \gg 1/L$ where
$L$ is the lattice length, to obtain the separation of scales shown in Fig~\ref{fig:scale1}.

Thus we are forced to consider masses that are sufficiently large
to satisfy this constraint.
Unfortunately, if we use too large a mass we will move the
system a long way from any IRFP and the
simple MCRG techniques we are using will not apply. The only obvious way
to reconcile these two things is
to use large enough boxes that we can keep $L \gg 1/m$ while making $m$
small enough to keep us close enough to
the CFT fixed point.

For our L=16 lattices to achieve for example $mL > 2$ we would need $am\gtrsim0.12$, which
may be simply too big to keep close the system close to the fixed point.

\section{Stability Matrix Method}
The two--lattice matching technique used in this work was first used to investigate quenched QCD~\cite{Hasenfratz:1984bx,Bowler:1984hv,Hasenfratz:1984hx}, and more recently QCD with many flavours of fermions~\cite{Hasenfratz:2009ea,Hasenfratz:2010fi}, and allows a determination of the most relevant coupling in a system. The original MCRG method~\cite{Swendsen:1979gn} in principle allows the extraction of all critical exponents of a system, both relevant and irrelevant.

\subsection{Method}
Consider a hamiltonian that can be written as a sum of couplings $K_i$ and observables $S_i$,
\begin{equation}
H = \sum_i K_iS_i,
\end{equation}
and an RG transform $R_s$ of scale $s$ such that
\begin{equation}
H^{(n+1)} = R_s H^{(n)} = \sum_i K^{(n+1)}_iS^{(n+1)}_i,
\end{equation}
where $S^{(n+1)}_i$ is the same observable as $S^{(n)}_i$ only measured on the lattice blocked $n+1$ rather than $n$ times. The fixed point of the RG transform is defined by the condition
\begin{equation}
H^{*} = R_s H^{*} = \sum_i K^{*}_i S^{*}_i,
\end{equation}
and near this point the flow in the couplings can be expanded linearly to give
\begin{equation}
K^{(n+1)}_i - K^{*}_i = \sum_jT_{ij}^{*}(K^{(n)}_j - K^{*}_j),
\end{equation}
where
\begin{equation}
T_{ij}^{*} = \left.\frac{\partial K^{(n+1)}_i}{\partial K^{(n)}_j}\right|_{H^{*}}.
\end{equation}

The chain rule gives
\begin{equation}
\frac{\partial \langle S^{(n)}_i\rangle}{\partial K^{(n-1)}_j} = \sum_k \frac{\partial K^{(n)}_k}{\partial K^{(n-1)}_j} \frac{\partial \langle S^{(n)}_i\rangle}{\partial K^{(n)}_k} = \sum_k T_{kj} \frac{\partial \langle S^{(n)}_i\rangle}{\partial K^{(n)}_k}
\end{equation}
From which $T_{kj}$ can be constructed using the identities
\begin{equation}
\frac{\partial \langle S^{(n)}_i\rangle}{\partial K^{(n-1)}_j} = \langle S^{(n)}_i S^{(n-1)}_j\rangle - \langle S^{(n)}_i\rangle\langle S^{(n-1)}_j\rangle \equiv A^{(n)}_{ij}
\end{equation}
\begin{equation}
\frac{\partial \langle S^{(n)}_i\rangle}{\partial K^{(n)}_j} = \langle S^{(n)}_i S^{(n)}_j\rangle - \langle S^{(n)}_i\rangle\langle S^{(n)}_j\rangle \equiv B^{(n)}_{ij}
\end{equation}

The eigenvalues of $T_{ij}^{*}$ give the critical exponents of the system~\cite{Wilson:1973jj}, e.g. $\nu = \ln s / \ln \lambda_h$, where $\lambda_h$ is the largest eigenvalue, and so $y_m = 1/\nu = \ln \lambda_h / \ln s$. From a single simulation close to the critical point, correlation functions of blocked observables are measured to construct the matrix $T_{ij}$, from which $y_m$ and other exponents can be determined.

If we are sufficiently close to a fixed point, then the largest eigenvalue of T should stay constant as the number of blocking steps is varied, and also as the number of observables used to construct T is varied. This method requires a larger lattice and higher statistics than the two--lattice method, but potentially allows more information to be extracted, in addition to being a useful consistency check of the two--lattice method results.

\subsection{Pure Gauge Results}
We have seven blocked observables and four blocking steps on the $32^4$ lattices. This means we can vary the number of observables, and hence the size of the stability matrix $T$, from 1 to 7. We can calculate $T$ after 1/2, 2/3 and 3/4 blocking steps, for any choice of our blocking parameter $\alpha$. Unlike in the two--lattice method there is no cancellation of finite size effects, so these are likely to be large.

We applied this method to our pure gauge configurations, where we expect a marginal largest eigenvalue of 1. Using more than four observables (i.e. including 8-link loops) tends to give a complex largest eigenvalue of T, so we only use 1 to 4 observables to construct T. Fig.~\ref{fig:Tpure} shows the largest eigenvalue after 1/2 (red), 2/3 (green) and 3/4 (blue) blocking steps, as a function of alpha using ORIG blocking, on a $32^4$ lattice at $\beta=3.0$:

\begin{figure}[ht]
  \centering
\includegraphics[angle=270,width=10cm]{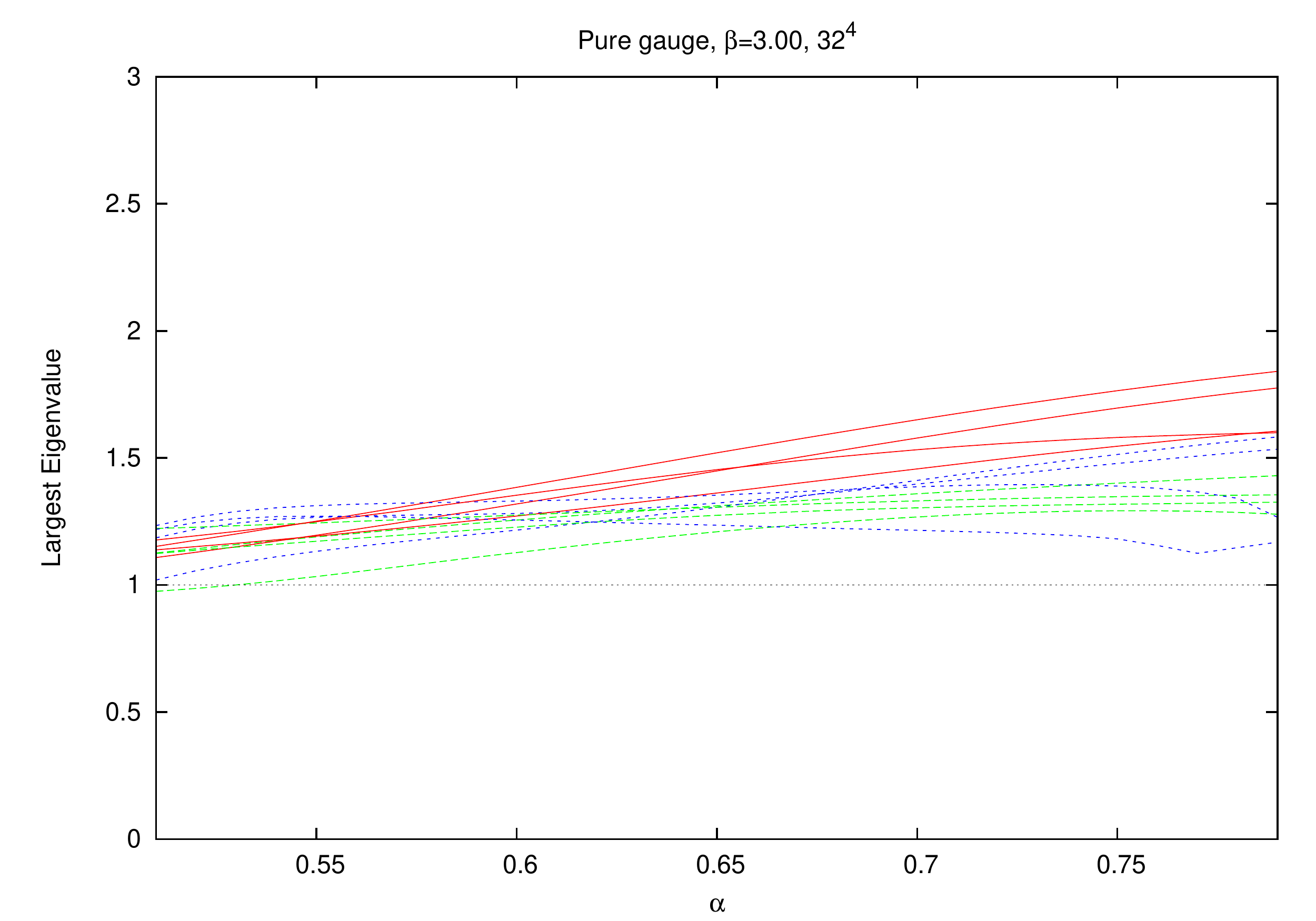}
\caption{The largest eigenvalue of the stability matrix T as a function of $\alpha$, after 1/2 (red), 2/3 (green) and 3/4 (blue) ORIG blocking steps on a $32^4$ pure gauge lattice at $\beta=3.0$. At each blocking step T is constructed using 1,2,3 and 4 observables, so the spread of eigenvalues of the same colour (same blocking level) gives some idea of the systematic variation from the number of observables included.}
  \label{fig:Tpure}
\end{figure}

The finite size effects are presumably very large, but in general this looks sensible - the variation with alpha is reduced as the number of blocking steps are increased, and for small $\alpha$ the eigenvalues are independent of the number of blocking steps within the spread of eigenvalues from varying the number of observables, and are consistent with a marginal eigenvalue of 1 corresponding to the expected logarithmic flow of the coupling in quenched Yang-Mills.

\subsection{MWT Results}

\begin{figure}[ht]
  \centering
\includegraphics[angle=270,width=10cm]{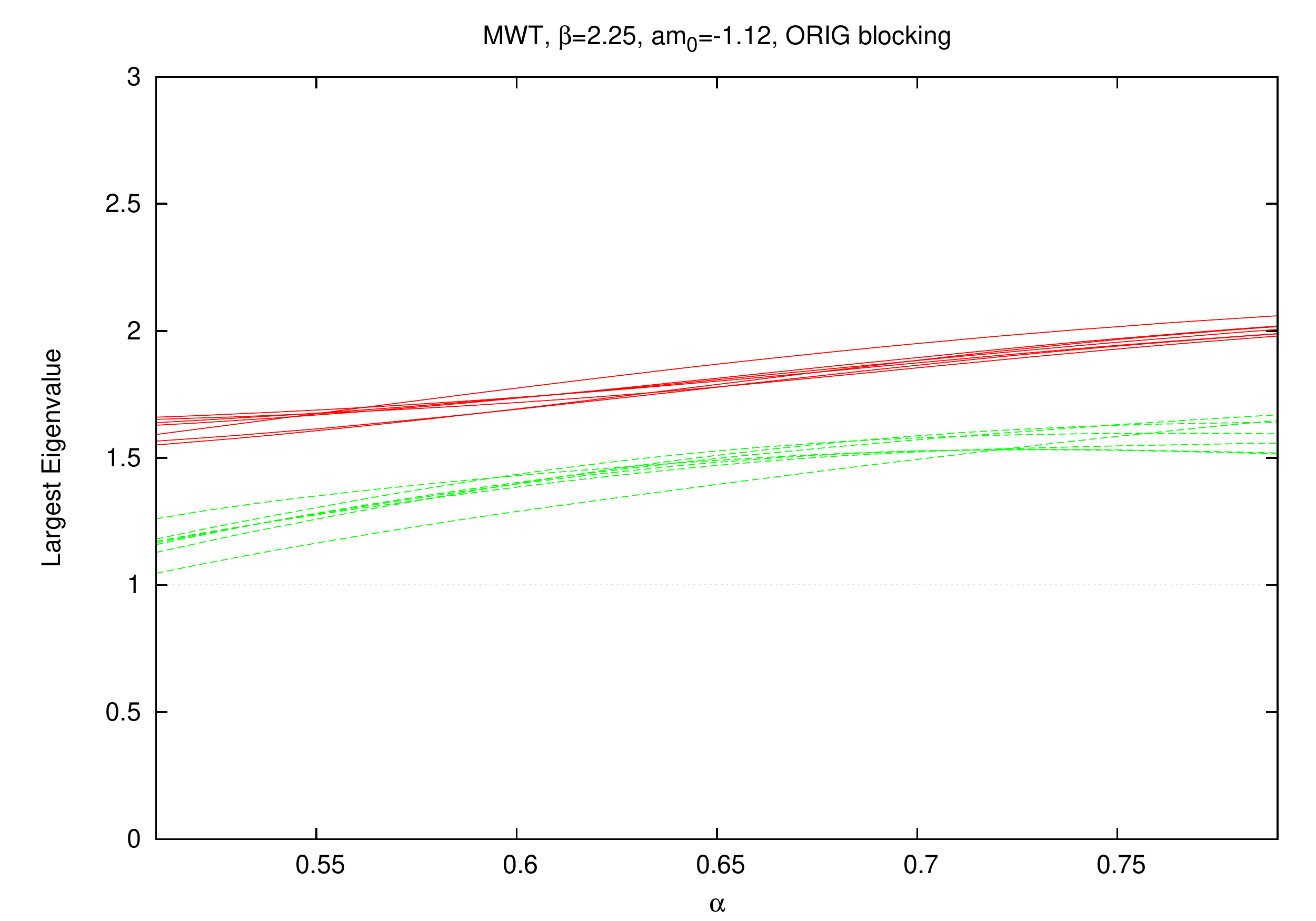}
\caption{The largest eigenvalue of the stability matrix T as a function of $\alpha$, after 1/2 (red) and 2/3 (green)  ORIG blocking steps on a $16^4$ lattice at $\beta=2.25$, $am\simeq0.2$. At each blocking step T is constructed using between 1 and 7 observables, so the spread gives some idea of the systematic variation from the number of observables included.}
  \label{fig:Tmwt}
\end{figure}

For Minimal Walking Technicolor we have $16^4$ lattices so we can construct T after 1/2 and 2/3 blocking steps, although again the finite size effects are likely to be large. Fig.~\ref{fig:Tmwt} shows the largest eigenvalue of T after 1/2 (red) and 2/3 (green) ORIG blocking steps for $\beta=2.25$, $am\simeq0.2$. Between 1 and 7 observables are used to construct T, and the spread of eigenvalues at a given blocking level is small, showing little dependence on the number of observables used. On the other hand there is a large difference between the two blocking steps, which suggests that we are not close to a fixed point.

This is representative of the situtation for all of our runs - the picture is qualitatively the same for virtually all of our values of $\beta$ and $m$, and for all three RG blocking transforms.

These results suggest that while the pure gauge simulations are sufficiently close to the renormalised trajectory, the MWT simulations are not close enough, and more blocking steps, and hence larger lattices, are required.

\FloatBarrier

\section{Conclusion}
We find a vanishing anomalous dimension and slow running of the coupling consistent with an IRFP, however our results suffer from considerable systematic errors. The most significant of these is the uncertainty in the location of the fixed point in the coupling, which in turn produces a large uncertainty in the value of the anomalous mass dimension. This is essentially the same issue that the Schr\"odinger Functional studies of this theory have suffered from.

Assuming negligible running in the coupling we find $\gamma = -0.03(13)$.
Using our errors in the
measurement of the discrete beta function to suggest a window for
the possible flow in beta produces a large uncertainty in the associated
anomalous dimension $-0.6 \lesssim \gamma \lesssim 0.6$. Previous 
studies~\cite{Bursa:2009we,Hietanen:2009az,DeGrand:2011qd,Giedt:2011kz} have provided evidence of a backward
flow in the coupling at these values of $\beta$ consistent with an IRFP.
If this is the case our results indicate
that the estimate of $\gamma=-0.03(13)$ assuming no flow in $\beta$
would actually constitute a {\it lower}
bound on the true anomalous dimension. A small value of $\gamma$ is also
likely to result from finite volume effects; if the beta function
is small but non--zero then many
blocking steps and hence large starting volumes
will be needed to move the system away from the usual
perturbative UV fixed point.

While the MCRG method is potentially a promising technique for studying possible conformal systems our simulations 
indicate that it is currently limited by several sources of systematic error. Perhaps the single largest factor contributing to this error is the relatively small lattices that have been used in this study; this places constraints on how small a quark mass can be used and limits the number of blocking steps that can be taken. These restrictions mean that in all likelihood the RG trajectory does not closely approach the putative IRFP making extraction of an exponent like $\gamma$ problematic.

Adding more matching observables, in particular fermionic ones such as meson correlators, should give a broader set of observables which will enable us to determine unambiguously matched actions. Using $32^4$ and $16^4$ lattices matching could be done after 1/2 and 2/3 blocking steps - the smallest blocked lattice being of size $4^4$, so that for example the meson correlator at each timeslice could be used as an observable.

Going to $32^4$ lattices will also allow us to go to smaller masses and thus be closer to the IRFP, as well as allowing us to match our current set of blocked observables after three blocking steps instead of two, which is a much more stringent matching condition that should help to constrain $\beta'$. With larger lattices we will also be able to check for finite volume effects using the analysis of Ref.~\cite{Hasenfratz:2011xn}. In addition it may be possible to get a determination of the anomalous mass dimension using the stability matrix MCRG method. Other potential improvements include using an improved action to reduce ${\mathcal O}(a)$ effects, and adding an adjoint plaquette term to the action to move the system away from a non--physical ultraviolet fixed point due to lattice artefacts, and allow us to go to stronger coupling~\cite{Hasenfratz:2011xn}.

However, even taking into account our currently large systematic errors, we find a small anomalous dimension which is clearly less than 1, and hence that, in its simplest form~\cite{Fukano:2010yv}, the SU(2) gauge theory with two Dirac fermions in the adjoint representation is not a viable Walking Technicolor candidate.

\section*{Acknowledgements}
JG is supported in part by the Department of Energy, Office of Science, Office of High Energy Physics under Grant No. DE-FG02-08ER41575. SMC is supported in part by DOE under Grant No. DE-FG02-85ER40237. We gratefully acknowledge the use of USQCD computing facilities on the lattice QCD cluster at Fermilab.

\appendix

\section{HYP Smearing}
\label{app:hyp}
A HYP smeared link $W[U]_{x,\mu}$ is constructed in three stages~\cite{Hasenfratz:2001hp}, each stage consisting of a modified APE blocking step~\cite{Albanese:1987ds} projected back into the gauge group. In the first stage a set of decorated links $\bar{V}$ are constructed from the original links $U$, 
\begin{equation}
\bar{V}_{n,\mu ;\nu \, \rho }=Proj\left[(1-\alpha _{3})U_{n,\mu }+\frac{\alpha _{3}}{2}\sum _{\pm \eta \neq \rho ,\nu ,\mu }U_{n,\eta }U_{n+\hat{\eta },\mu }U_{n+\hat{\mu },\eta }^{\dagger }\right].
\end{equation}
From these a second set of decorated links $\tilde{V}$ are constructed,
\begin{equation}
\tilde{V}_{n,\mu ;\nu }=Proj\left[(1-\alpha _{2})U_{n,\mu }+\frac{\alpha _{2}}{4}\sum _{\pm \rho \neq \nu ,\mu }\bar{V}_{n,\rho ;\nu \, \mu }\bar{V}_{n+\hat{\rho },\mu ;\rho \, \nu }\bar{V}_{n+\hat{\mu },\rho ;\nu \, \mu }^{\dagger }\right],
\end{equation}
and finally the HYP smeared links are constructed from this second set of dressed links,
\begin{equation}
W[U]_{n,\mu}=Proj\left[(1-\alpha _{1})U_{n,\mu }+\frac{\alpha _{1}}{6}\sum _{\pm \nu \neq \mu }\tilde{V}_{n,\nu ;\mu }\tilde{V}_{n+\hat{\nu },\mu ;\nu }\tilde{V}_{n+\hat{\mu },\nu ;\mu }^{\dagger }\right].
\end{equation}

\section{Observables}
\label{app:obs}
Explicit expressions for the seven observables used in the matching: 
\begin{equation}
\begin{array}{lll}
{\mathcal O}_1 & = & \sum_{\mu=1}^{4}\sum_{\nu=0}^{\mu} P(\mu,\nu) \\
{\mathcal O}_2 & = & \sum_{\mu=1}^{4}\sum_{\nu=0}^{\mu} L_6(\mu,\nu,\rho=(\mu+1)\bmod{4}) \\
{\mathcal O}_3 & = & \sum_{\mu=1}^{4}\sum_{\nu=0}^{\mu} L_6(\mu,\nu,\rho=(\mu+2)\bmod{4}) \\
{\mathcal O}_4 & = & \sum_{\mu=1}^{4}\sum_{\nu=0}^{\mu} L_6(\mu,\nu,\rho=(\mu+3)\bmod{4}) \\ 
{\mathcal O}_5 & = & \sum_{\mu=1}^{4}\sum_{\nu=0}^{\mu} L_8(\mu,\nu,\rho=(\mu+1)\bmod{4},\alpha=(\mu+1)\bmod{4}) \\
{\mathcal O}_6 & = & \sum_{\mu=1}^{4}\sum_{\nu=0}^{\mu} L_8(\mu,\nu,\rho=(\mu+2)\bmod{4},\alpha=(\mu+1)\bmod{4}) \\
{\mathcal O}_7 & = & \sum_{\mu=1}^{4}\sum_{\nu=0}^{\mu} L_8(\mu,\nu,\rho=(\mu+3)\bmod{4},\alpha=(\mu+1)\bmod{4}) \\
\end{array}
\end{equation}
where
\begin{equation}
\begin{array}{lll}
P(\mu,\nu) & = & \sum_x\Re\Tr \left[U_{x,\mu}U_{x+\mu,\nu}U^{\dagger}_{x+\nu,\mu}U^{\dagger}_{x,\nu}\right] \\
L_6(\mu,\nu,\rho) & = & \sum_x\Re\Tr \left[U_{x,\mu}U_{x+\mu,\nu}U_{x+\mu+\nu,\rho}U^{\dagger}_{x+\nu+\rho,\mu}U^{\dagger}_{x+\nu,\rho}U^{\dagger}_{x,\nu}\right] \\
L_8(\mu,\nu,\rho) & = & \sum_x\Re\Tr \left[U_{x,\mu}U_{x+\mu,\nu}U_{x+\mu+\nu,\rho}U_{x+\mu+\nu+\rho,\alpha}U^{\dagger}_{x+\nu+\rho+\alpha,\mu}U^{\dagger}_{x+\nu+\rho,\alpha}U^{\dagger}_{x+\nu,\rho}U^{\dagger}_{x,\nu}\right] \\
\end{array}
\end{equation}

\section{PCAC masses}
\label{app:pcac}
\begin{figure}[ht]
  \centering
\includegraphics[angle=270,width=12.0cm]{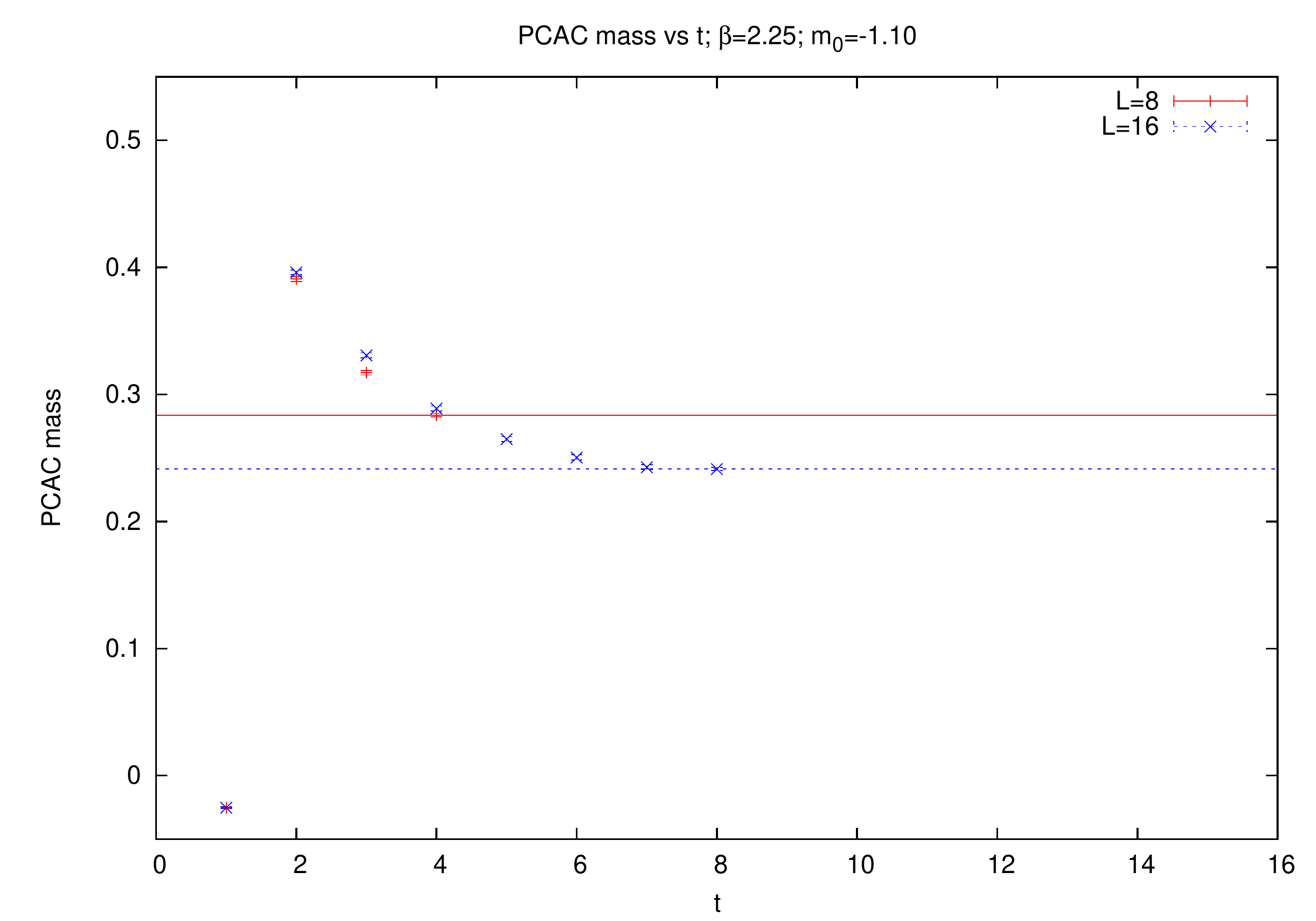}

\includegraphics[angle=270,width=12.0cm]{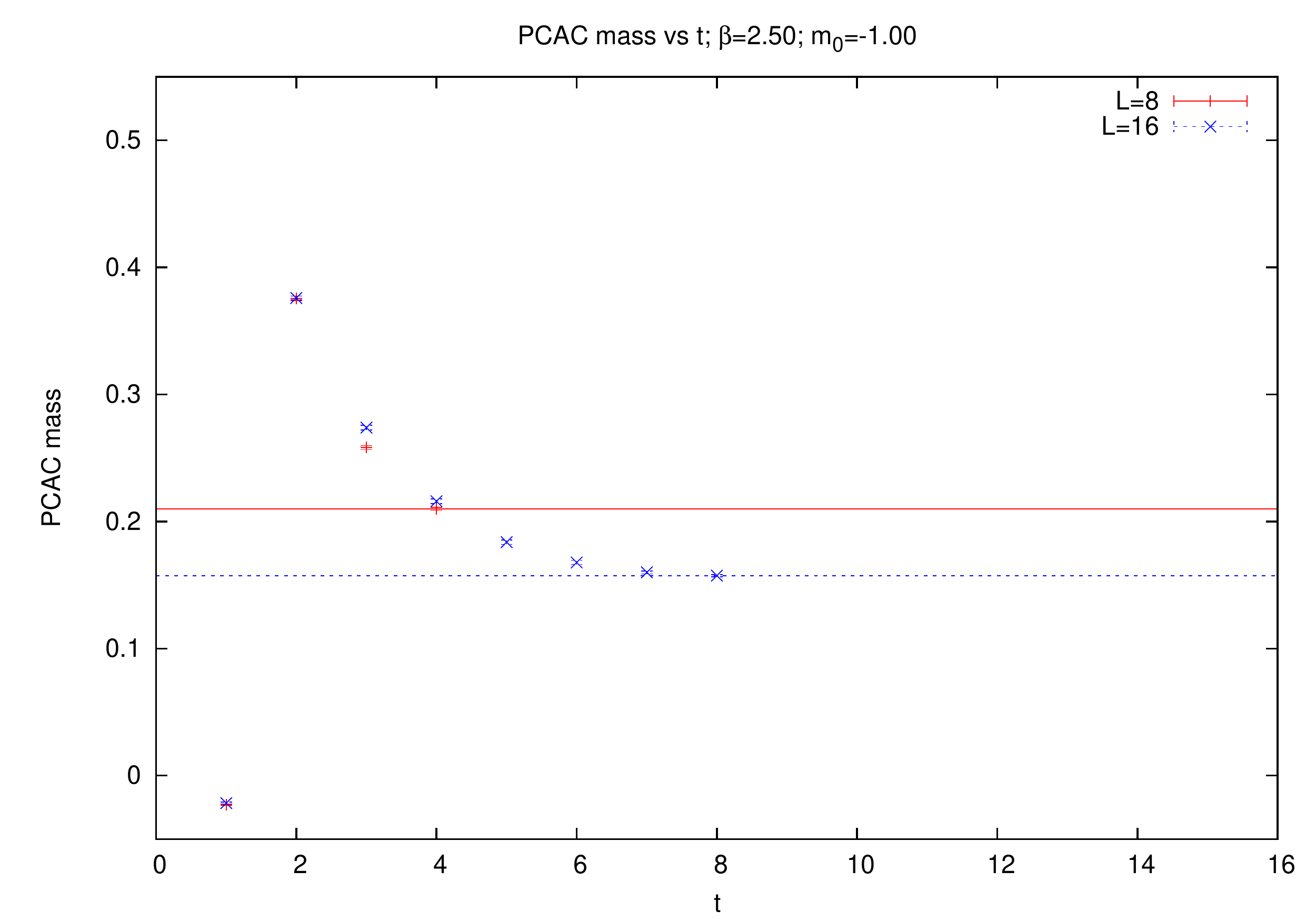}  
\caption{PCAC mass as a function of time on $8^4$ and $16^4$ lattices.}
  \label{fig:MASS_PCACt}
\end{figure}

The PCAC mass measured on the $8^4$ lattices suffers from a finite--volume effect, the time extent of the lattice is not sufficient for the mass to reach a plateau. This can be seen in Fig.~\ref{fig:MASS_PCACt}, where the PCAC mass for the same values of $(\beta,am)$ on lattices of size $8^4$ and $16^4$ are compared.

\begin{figure}[ht]
  \centering
\includegraphics[angle=270,width=12.0cm]{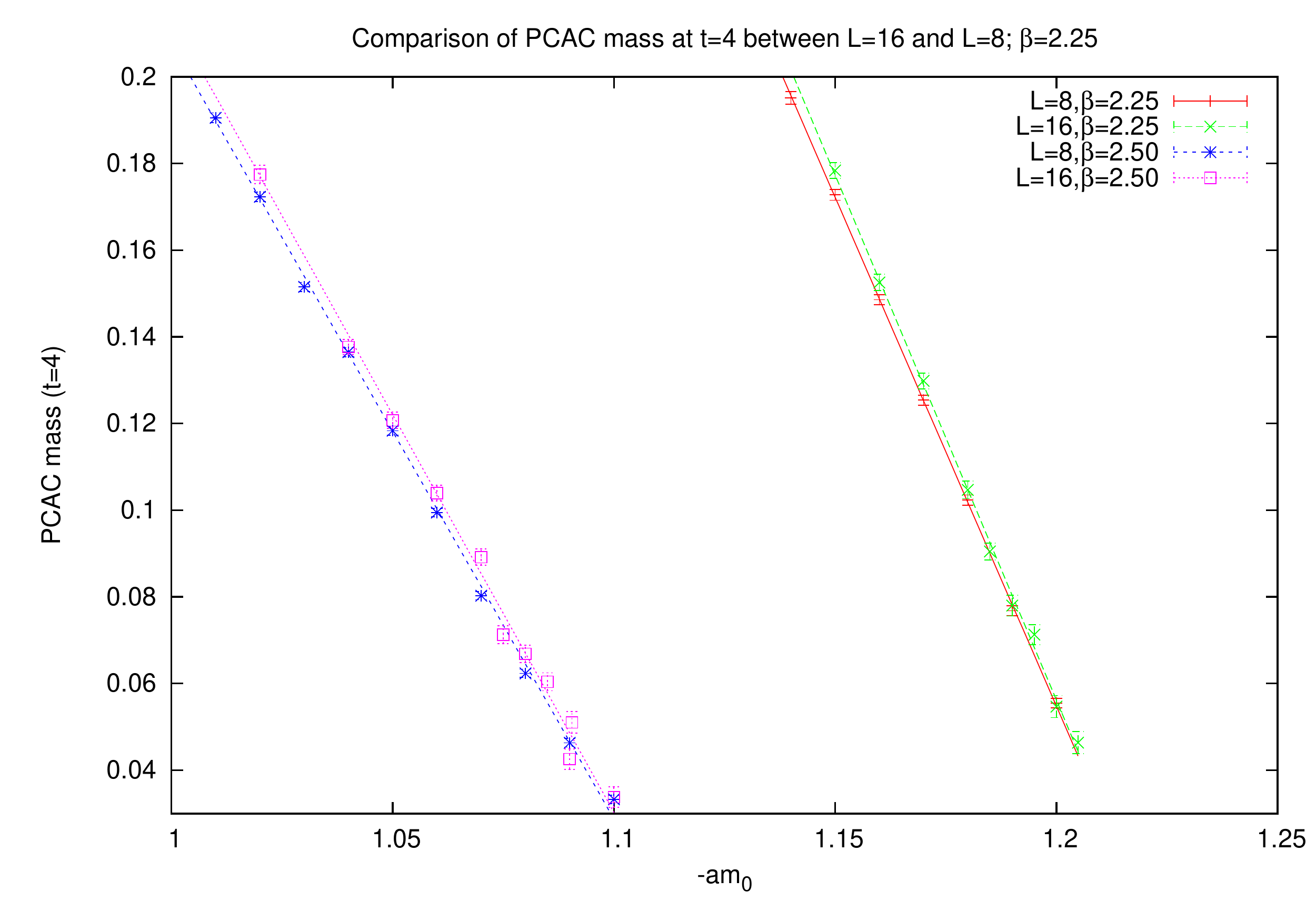}  
\caption{PCAC mass at t=4 on $8^4$ and $16^4$ lattices.}
  \label{fig:MASS_PCACt4}
\end{figure}

The lattice artefacts however seem to be small - the PCAC mass as a function of $t$ agrees well between the two lattices. We use the PCAC mass measured at $t=8$ on the $16^4$ lattices throughout this paper to convert bare masses to PCAC masses. To estimate the systematic error due to this procedure, we can compare the PCAC mass at the same timeslice $t=4$ on the two lattice sizes, as shown in Fig.~\ref{fig:MASS_PCACt4}. The difference is smaller than the statistical error on the PCAC mass for small masses, and increases approximately linearly with mass to $\sim0.005$ at $am=0.20$. To include this source of error, a systematic error of $0.025am$ was added linearly to the measured error on each matching mass.

\section{Matching mass range}
\label{app:mass}
We have assumed that the contribution from the irrelevant directions is negligible by the time we perform the matching. But in reality we match after only one or two steps, and there may still be a contribution coming from the coupling. We are forcing the matching coupling to the same value as the initial coupling. This causes a shift in the measured observable, which is then cancelled by a shift in the observed matching mass from the true value.

Consider the situation where we measure some observable ${\mathcal O}^{(n)}(\beta,am)$ on the larger lattice after $n$ blocking steps. The correct matching done in both the coupling and the mass would give $a'm',\beta'$ such that
\[{\mathcal O}^{(n-1)}(\beta',a'm') = {\mathcal O}^{(n)}(\beta,am)\]

Instead we fix $\beta'=\beta$ and find $a''m''$ such that
\[{\mathcal O}^{(n-1)}(\beta,a''m'') = {\mathcal O}^{(n)}(\beta,am)\]

Dropping the $^{(n-1)}$ superscript on ${\mathcal O}$ and taylor expanding around the correct matching values gives

\[{\mathcal O}(\beta,a''m'') = {\mathcal O}(\beta',a'm') + c_{\beta}(\beta-\beta') + c_m(a''m''-a'm')\]
where
\[c_{\beta} = \left.\frac{\partial {\mathcal O}}{\partial \beta}\right|_{\beta',a'm'}, \quad c_m = \left.\frac{\partial {\mathcal O}}{\partial a''m''}\right|_{\beta',a'm'}\]
This gives the relation
\[a''m'' = a'm' -\frac{c_{\beta}}{c_m}(\beta-\beta').\]
So the effect of not matching in $\beta'$ will cause the measured $a''m''$ matched to $am=0$ to be shifted away from zero.

For this reason we excluded the matchings at $am<0.02$ from the anomalous mass dimension fits, since they are likely to have significant contributions from the running of the coupling, although including them makes little difference to the fits.

In fact the size of this shift at $am=0$ could give us an indirect way to measure the running of the coupling. At $am=0$, we know that $a'm'=0$, which gives the relation
\[(\beta-\beta') = -\frac{c_m}{c_{\beta}}a''m''.\]
The difficulty in determining $c_m$ and $c_{\beta}$, as well as the relatively large uncertainty on the measurement of $a''m''$, means that although in principle this would be a way to measure $(\beta-\beta')$, in practice it is not feasible with our data. 

\section{Massless interpolation}
\label{app:massless}
\begin{figure}[ht]
  \centering
\includegraphics[angle=270,width=12.0cm]{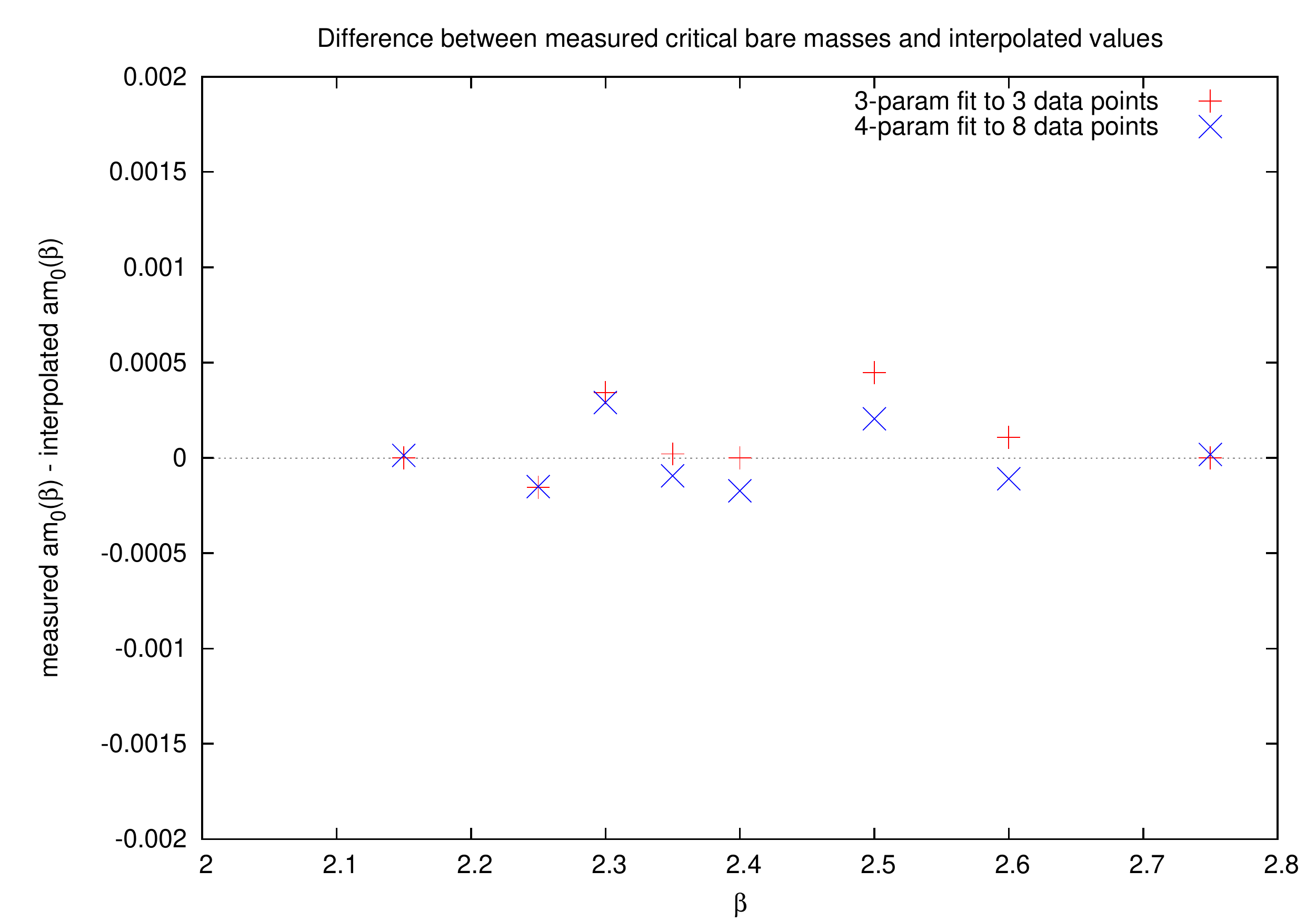}
  \caption{To estimate the systematic error, here is the difference between measured critical bare masses, and the values given by a 3--parameter interpolation function fitted to only three of the measured values. Also shown is the difference between measured values and a 4--parameter interpolation function fit to all the data. The difference in PCAC mass is $(1.5-3.5)\times$ this bare mass difference, which is still $\lesssim 0.001$, and smaller than the statistical uncertainty on the measured PCAC mass.}
  \label{fig:MASSLESS_fit}
\end{figure}

The difference between the measured critical bare masses, and the values predicted by the interpolation function, are very small as shown in Fig.~\ref{fig:MASSLESS_fit}. This implies a systematic error $\lesssim 0.001$ on the PCAC masses due to the interpolating function.

\bibliographystyle{unsrt}

\end{document}